\def\1{_{\vert}}
\font\bigastfont=cmr10 scaled \magstep 3
\def\bdot{\hbox{\bigastfont .}}
\def\ueber#1#2{{\setbox0=\hbox{$#1$}%
  \setbox1=\hbox to\wd0{\hss$\scriptscriptstyle #2$\hss}%
  \offinterlineskip
  \vbox{\box1\kern0.4mm\box0}}{}}
\def\ref{\par\noindent\hangindent\parindent\hangafter1}
\documentstyle[12pt,psfig]{ioplppt}          
\begin{document}
\title{Lagrangian Perturbation Approach \\ to the Formation of Large--scale
Structure$^*$}

\author{Thomas Buchert}

\address{Theoretische Physik, Ludwig--Maximilians--Universit\"at\\ 
Theresienstr. 37, D--80333 M\"unchen, Germany}

\begin{abstract}
The present lecture notes address three columns on which the Lagrangian
perturbation approach to cosmological dynamics is based: 1. the formulation
of a Lagrangian theory of self--gravitating flows in which the dynamics is 
described in terms of a single field variable; 
2. the procedure, how to obtain 
the dynamics of Eulerian fields from the Lagrangian picture, 
and 3. a precise definition of 
a Newtonian cosmology framework in which Lagrangian perturbation solutions 
can be studied.
While the first is a discussion of the basic equations obtained by transforming
the Eulerian evolution and field equations to the Lagrangian picture, 
the second exemplifies how the Lagrangian  
theory determines the evolution of Eulerian fields 
including kinematical variables like
expansion, vorticity, as well as the shear and tidal tensors. 
The third column is based on a specification of initial and boundary conditions,
and in particular on the identification of the average flow 
of an inhomogeneous 
cosmology with a ``Hubble--flow''. Here, we also look at the limits of the 
Lagrangian perturbation  
approach as inferred from comparisons with N--body simulations and illustrate some
striking properties of the solutions.
\end{abstract}

\vskip -0.5 true cm

{$^{*}$ \footnotesize to appear in: Proc.~Int. School of Physics
Enrico Fermi, Course CXXXII, Varenna 1995.}

\section{Lagrangian Theory of Self--gravitating Flows}

The description of fluid motions in cosmology has been largely studied 
in an {\it Eulerian} coordinate system $\vec x$, i.e., a rectangular non--rotating frame in
Euclidean space. Quite recently, it has become popular to study fluid motions
in a {\it Lagrangian} coordinate system $\vec X$, i.e., a curvilinear, possibly 
rotating frame in Euclidean space which is defined such as to move with the 
fluid. Since the Lagrangian description has a number of advantages over
the Eulerian one, and since this description enjoys many applications in
the recent cosmology literature, it is important to elucidate in proper language the
Lagrangian formalism.
Since the lectures by Fran\c cois Bouchet and Peter Coles (this volume) 
explore the field of recent applications of the Lagrangian perturbation theory,
I here concentrate on the basic architecture of a Lagrangian theory of 
structure formation. I do this in Newtonian cosmology, the lecture by Sabino
Matarrese (this volume) gives an extension to 
the framework of General Relativity.
Accordingly, I kept my reference list short, since more references may be found
in the other lectures.

That the Lagrangian approach is experiencing a revival in cosmology is good news;
I consider it the natural frame to describe fluid motions, since this description is 
formally close to the mechanics of point particles.
If you consult old textbooks on hydrodynamics, you will find that the 
Lagrangian picture was considered too complicated for practical purposes beyond
problems with high symmetry, and therefore has not been pursued further.
I hope that, after this and the related lectures, 
you will be convinced of the opposite.
   
\bigskip

Let us start with the basic system of equations in Newtonian
cosmology describing the motion of a pressureless fluid in the gravitational field 
which is generated by its own density. We think at applications for a 
(dominating) collisionless component in the Universe; the 
gravitational dynamics we describe is thought to act as an attractor for the baryonic matter
component which is ``lighted up'' by physics not described by these equations.

With this assumption the fluid motion in Eulerian space is completely characterized
by its velocity field $\vec v (\vec x,t)$ 
and its density field $\varrho(\vec x,t) >0$. The fluid has to obey the familiar 
{\it evolution equations} for these fields,
$$
\eqalignno{&\partial_t \vec v = - (\vec v \cdot \nabla) \vec v + \vec g  \; ,
&(1a)\cr
&\partial_t \varrho = - \nabla \cdot(\varrho \vec v) \; ,
&(1b)\cr}
$$
where the gravitational field $\vec g (\vec x,t)$ is constrained by the 
(Newtonian) {\it field equations}
$$
\eqalignno{
&\nabla \times \vec g = \vec 0 \; , &(1c) \cr
&\nabla \cdot \vec g = \Lambda - 4 \pi G \rho \; ; &(1d) \cr}
$$
$\Lambda$ denotes the cosmological constant.
(Strictly speaking, $\vec g$ in eq. (1a) is a force per unit
{\it inertial mass}, whereas in eqs. (1c,d) $\vec g$ is the field strength
associated with {\it gravitational mass}. That we set both equal is the content
of Einstein's equivalence principle of inertial and gravitational mass.)

Alternatively, we can write the eqs. (1c) and (1d) in terms of a single 
Poisson equation for the gravitational potential, which we do not need in 
the following. Hereafter, we call the system (1) 
the {\it Euler--Newton system}.

One important issue to learn about the Lagrangian treatment is 
the fact that both evolution equations (1a) and (1b) can be integrated 
exactly in Lagrangian space, velocity and density will therefore not appear
as dynamical variables later.
To see this, we first look at the basic Lagrangian field variable which is the
{\it trajectory field} of fluid elements, or the {\it deformation field} of 
the medium, respectively (Fig.1a):
$$
\vec x = {\vec f}(\vec X,t) \;\;;\;\;\vec X : = {\vec f}(\vec X,t_0)\;\,,
\eqno(2a)
$$
where $\vec X$ denote the Lagrangian coordinates which label fluid elements,
$\vec x$ are the positions of these elements in Eulerian space at the
time $t$, and $\vec f$ is the trajectory of fluid elements for
constant $\vec X$. (Notice: the Eulerian positions $\vec x$ are here viewed not as independent
variables (i.e., coordinates), 
but as dependent fields of Lagrangian coordinates; therefore, we employ the letter $\vec f$
for the sake of clarity. Independent variables
are now $(\vec X,t)$ instead of $(\vec x,t)$.)

\begin{figure} 

\psfig{figure=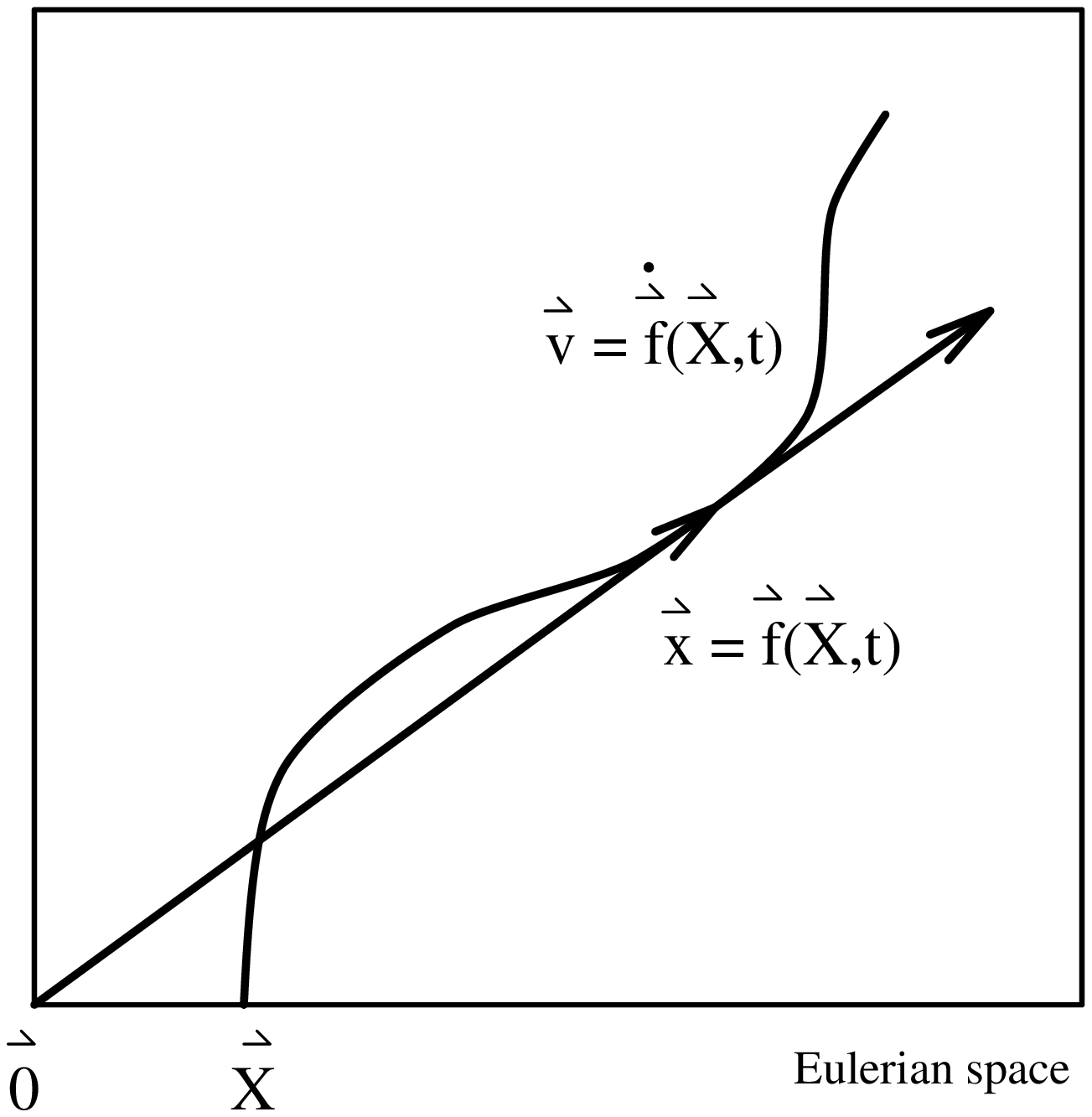,height=7.0cm,width=7.0cm}
\vskip -7.1 true cm
\rightline{\psfig{figure=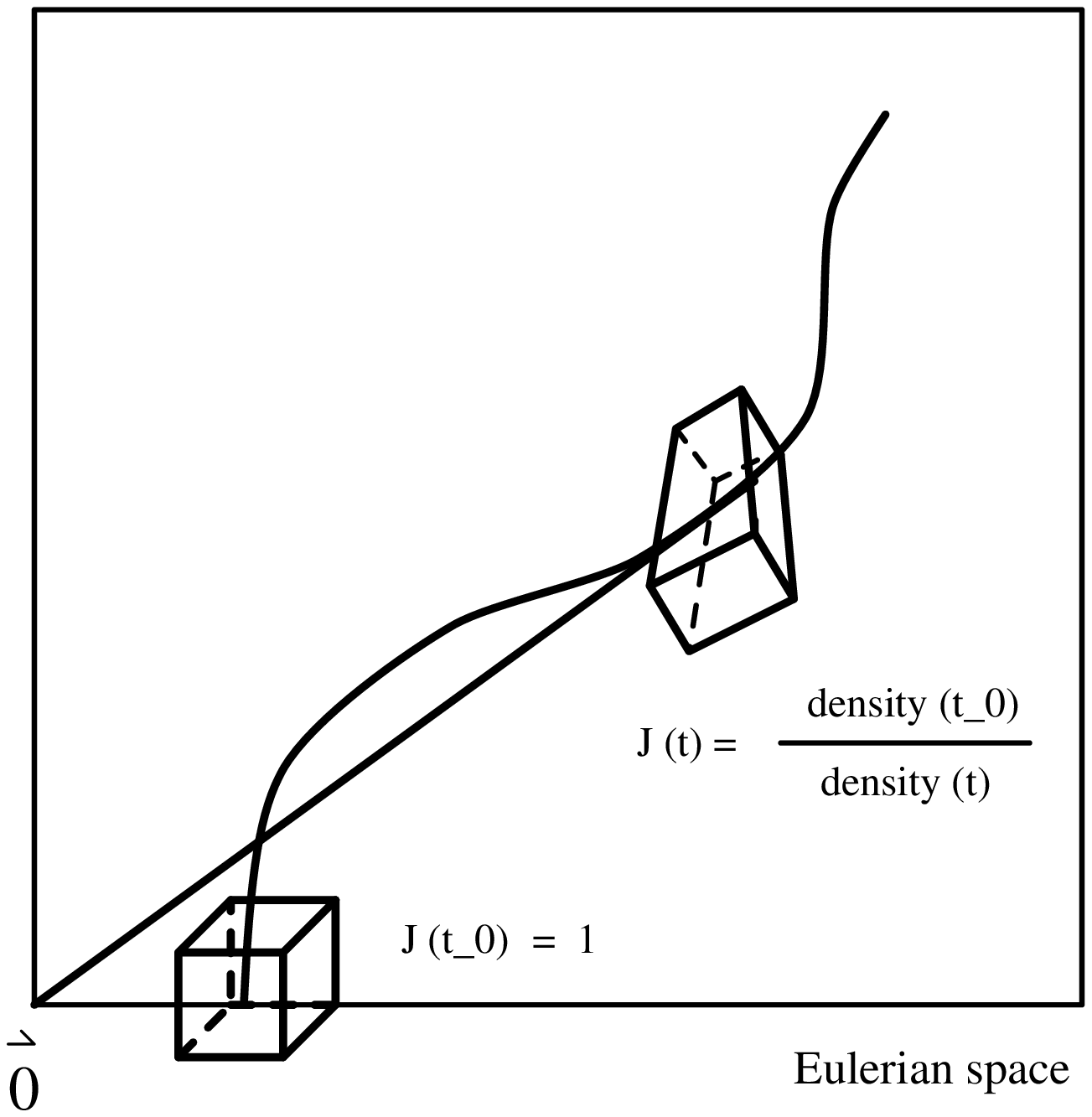,height=7.0cm,width=7.0cm}}

\caption{
The Newtonian spacetime projected onto an Eulerian plane
is scetched. A fluid element sitting initially ($t=t_0$)
at $\vec X$ moves along the trajectory $\vec f$ to the Eulerian position
$\vec x$ at time $t$; this position vector is expressed in terms of Lagrangian
(i.e. initial) coordinates which are constant along the 
trajectory: ${\dot{\vec X}} = \vec 0$ (Fig.1a). The Jacobian $J$ measures the volume
deformation of fluid elements located at $\vec X$; $J$ drops from $1$ to zero as the 
volume element degenerates into a surface element (piece of pancake), a line element
(piece of filament), or a point (cluster).}

\end{figure}

\noindent
The velocity field $\vec v (\vec x,t)$ of the fluid is the
tangential
field to this family of curves, 
$$
{\vec v}=\dot {\vec f}(\vec X,t)\;\;,\eqno(2b)
$$
and the dot denotes
the total (Lagrangian or convective) time--derivative along 
$\vec v$:
$$
\left(...\right)^{\bdot}\equiv
{d\over dt} := 
\partial_t  + \vec v \cdot \nabla \;\;.\eqno(2c)
$$
(The dot commutes with Lagrangian differentiation.)

\noindent
In view of (2c) we recognize that the Eulerian evolution equation (1a) 
can be written as $\dot{\vec v}=\vec g$, which is the
familiar definition of acceleration (here, the acceleration of a fluid element along
its trajectory). Consequently, in view of (2b), 
we know the acceleration field in Lagrangian space 
for {\it any given} trajectory:
$$
\vec g = \ddot{\vec f}(\vec X,t)\;\;.
\eqno(2d)
$$
In other words, taking (2b) and (2d) as {\it definitions} of velocity and
acceleration, respectively (like in point mechanics!), we see that  
equation (1a) is automatically fulfilled as we switch
to the Lagrangian description.

A similar logic applies to the continuity equation (1b): 
the deformation of the medium is described by the Lagrangian {\it deformation
tensor} ($f_{i\1 k}$), i.e., the tensor of first derivatives of the trajectory 
field with respect to Lagrangian coordinates\ftnote{1}{Lagrangian 
differentiation is indicated throughout this lecture
with a vertical slash to make a difference to differentiation with respect to
Eulerian coordinates, denoted by commata.}
which measures how much the Eulerian 
positions deviate from their original (Lagrangian) positions.
The volume of the deformed fluid element is measured by the determinant of 
this tensor, $J(\vec X,t):=\det(f_{i\1 k})$, and therefore its density must be inversely
proportional to it (Fig.1b):
$$
\varrho (\vec X, t) = \ueber{\varrho}{o} J^{-1} \;\;.\eqno(2e)
$$  
($\ueber{\varrho}{o}:=\varrho (\vec X, t_0)$ is the initial density field, and
$J(\vec X,t_0)=1$ according to our definition of the Lagrangian coordinates
in (2a) which coincide with the Eulerian ones at $t=t_0$.)
You may verify (2e) by differentiation and by using the identity
$$
\dot J = J \nabla \cdot \vec v\;\;.\eqno(2f)
$$
Again we conclude that, for {\it any given} $\vec f (\vec X,t)$, we obtain
the exact expression for the density at the fluid element, i.e., in Lagrangian
space. 

Let us now move to the question which {\it Lagrangian evolution equations} the
field of trajectories has to obey. We obtain them by expressing the 
field equations (1c) and (1d) (which are $4$ linear ``constraint equations'' 
for the 
acceleration field) in terms of Lagrangian coordinates which, as we shall see,
yields $4$ non--linear partial differential 
equations for $\vec g (\vec X,t)$.
For this purpose we need the inverse of the transformation (2a) from Lagrangian
to Eulerian coordinates,
$$
\vec X = \vec h (\vec x,t)\;\;;\;\;\vec h \equiv {\vec f}^{-1}\;\;.
\eqno(2g)
$$
While the deformation tensor is the Jacobian matrix of the transformation
from Eulerian to Lagrangian coordinates, $f_{i\1 k} = J_{ik}$, the inverse
of $f_{i\1 k}$ is the inverse Jacobian matrix\ftnote{2}{Hereafter 
we adopt the summation convention.}
$$ 
h_{a,b} = J_{ab}^{-1} = ad(J_{ab})J^{-1} = {1\over 2J}\epsilon_{ajk}\epsilon_{b\ell m}
f_{\ell\1 j}f_{m\1 k}\;\;.\eqno(2h)
$$ 
Since the transformation part ``looks'' inconvenient during a first
reading of papers on Lagrangian models, 
I here try to be 
as elementary as possible. 

We look at the two--dimensional case first, i.e., there are  
only two non--vanishing field components, e.g., $g_1$ and $g_2$
\ftnote{1}{We are 
talking about a purely two--dimensional space and not about cylinders in the 
third direction in which case we would have $f_3 = X_3$.}.
We then get for the Eulerian derivatives of $\vec g$:
$$
\eqalignno{
g_{1,1}&=g_{1\1 1}h_{1,1}+g_{1\1 2}h_{2,1}\;\;,\;\;
g_{1,2}=g_{1\1 1}h_{1,2}+g_{1\1 2}h_{2,2}\;\;,\cr
g_{2,1}&=g_{2\1 1}h_{1,1}+g_{2\1 2}h_{2,1}\;\;,\;\;
g_{2,2}=g_{2\1 1}h_{1,2}+g_{2\1 2}h_{2,2}\;\;.
&(3a)\cr}
$$ 
Inserting the inverse Jacobian matrix,
$$
J_{ab}^{-1}(2D)=
{1\over J}\pmatrix{f_{2\1 2}&-f_{1\1 2}\cr-f_{2\1 1}&f_{1\1 1}\cr}\;\;,\eqno(3b)
$$
with $J(2D)=f_{1\1 1}f_{2\1 2} - f_{1\1 2}f_{2\1 1}$, 
we can express the single component of the 
curl and the divergence of $\vec g$ in terms of Lagrangian derivatives of
$\vec g(\vec X,t)$ and $\vec f(\vec X,t)$: 
$$
\eqalignno{
g_{2,1}-g_{1,2}&=\left(g_{2\1 1}f_{2\1 2} - g_{2\1 2}f_{2\1 1} + 
g_{1\1 1}f_{1\1 2} - g_{1\1 2}f_{1\1 1}\right)J^{-1}\;\;, 
&(4a)\cr
g_{1,1}+g_{2,2}&=\left(g_{1\1 1}f_{2\1 2} - g_{1\1 2}f_{2\1 1} - 
g_{2\1 1}f_{1\1 2} + g_{2\1 2}f_{1\1 1}\right)J^{-1}\;\;. 
&(4b)\cr}
$$
Using the exact integrals (2d) and (2e) of eqs. (1a) and (1b), eqs. (1c) and (1d) are 
transformed into a closed Newtonian system for the trajectories in 2D:
$$
\eqalignno{
&{\ddot f}_{2\1 1}f_{2\1 2} - {\ddot f}_{2\1 2}f_{2\1 1} + 
{\ddot f}_{1\1 1}f_{1\1 2} - {\ddot f}_{1\1 2}f_{1\1 1}\;=\;0\;\;,
&(5a)\cr
&{\ddot f}_{1\1 1}f_{2\1 2} - {\ddot f}_{1\1 2}f_{2\1 1} - 
{\ddot f}_{2\1 1}f_{1\1 2} + 
{\ddot f}_{2\1 2}f_{1\1 1}\;=\;\Lambda J(2D) - 4\pi G \ueber{\varrho}{o}\;\;.
&(5b)\cr}
$$
(Notice that we have multiplied with $J$, i.e., we must ensure that $J\ne 0$.)
Looking at the equations (5) we see that we have just written out functional 
determinants:
$$
\eqalignno{
&{\partial({\ddot f}_1,f_1)\over \partial(X_1,X_2)}+
{\partial({\ddot f}_2,f_2)\over \partial(X_1,X_2)}
\;=\;0\;\;,
&(5a)\cr
&{\partial({\ddot f}_1,f_2)\over \partial(X_1,X_2)}-
{\partial({\ddot f}_2,f_1)\over \partial(X_1,X_2)}
\;=\;\Lambda {\partial(f_1,f_2)\over \partial(X_1,X_2)} 
- 4\pi G \ueber{\varrho}{o}\;\;.
&(5b)\cr}
$$
In 3D we have to employ some more tensor algebra, but the procedure is the same. 
The relevant formula 
for the transformation of any tensor (here exemplified for the acceleration
gradient ($g_{i,j}$)) reads:
$$
g_{i,j}=g_{i\1 k}h_{k,j} = {1\over 2J} \epsilon_{k\ell m}\epsilon_{jpq}
g_{i\1 k} f_{p\1 \ell} f_{q\1 m}\;\;,
\eqno(6)
$$
where we have used the formula for the inverse Jacobian (2h).
In view of the definition of a functional determinant of any three functions
$A(\vec X,t)$, $B(\vec X,t)$ and $C(\vec X,t)$,
$$
{\cal J}(A,B,C):=
{\partial (A, B, C)\over
\partial(X_1,X_2,X_3)}=\epsilon_{k\ell m}A_{\1 k}B_{\1 \ell}C_{\1 m}\;\;,
$$
e.g., for the Jacobian determinant we simply have $J={\cal J}(f_1,f_2,f_3)$,
we can write equation (6) as 
$$
g_{i,j}= {1\over 2J} \epsilon_{jpq}{\cal J}(g_i,f_p,f_q)\;\;.
\eqno(6)
$$
The curl and the divergence of the acceleration field can be read 
from eq. (6) as 
the anti--symmetric part of the acceleration gradient and its trace 
(here, repeated indices imply
summation as before, but with $i,j,k$ running through the cyclic permutations of
$1,2,3$):
$$
\eqalignno{
&g_{[i,j]}=-{1\over 2}(\nabla \times \vec g)_k = {1\over 2}
\epsilon_{pq \lbrack j} {\cal J}({\ddot f}_{i
\rbrack},f_p,f_q) J^{-1}
\;\;,&(7a,b,c)\cr
&g_{i,i}=(\nabla \cdot \vec g) =
{1 \over 2}\epsilon_{abc}\;{\cal J}({\ddot f}_a,f_b,f_c) J^{-1}
\;\;.&(7d)\cr}
$$
Inserting (7) and the exact integrals (2d) and (2e) 
into (1c) and (1d) we finally obtain the 
{\it Lagrange--Newton system} 
$\lbrack 7 \rbrack$(no background source, in particular $\Lambda=0$) 
and $\lbrack 8 \rbrack$(including backgrounds of 
Friedmann type):
$$
\eqalignno{
{\cal J}({\ddot f}_1, f_1, f_3)&+{\cal J}({\ddot f}_2, f_2, f_3)
\;=0\;\;,
&(8a)\cr
{\cal J}({\ddot f}_2, f_2, f_1)&+{\cal J}({\ddot f}_3, f_3, f_1)
\;=0\;\;,
&(8b)\cr
{\cal J}({\ddot f}_1, f_1, f_2)&+{\cal J}({\ddot f}_3, f_3, f_2)
\;=0\;\;,
&(8c)\cr
{\cal J}({\ddot f}_1, f_2, f_3)&+{\cal J}({\ddot f}_2,f_3,f_1)
+{\cal J}({\ddot f}_3,f_1,f_2) \cr&
- \Lambda \;{\cal J}(f_1,f_2,f_3)\;
=-4\pi G \ueber{\varrho}{o}
\;\;.&(8d)\cr}
$$
I have written this system here in explicit form
to make working with these equations most
convenient. (Alternative forms of these equations may be found in 
$\lbrack 21 \rbrack$; in that paper we also employ the calculus of differential forms, 
which makes the above derivation even simpler.) 
\bigskip

The Lagrange--Newton system (8) is the basic system of equations we want 
to study. Although these equations look, and in fact are more complicated 
than their Eulerian counterparts, they have proven to be as useful for 
finding exact solutions as well as perturbative approximations beyond the 
linear regime.
Here, I note that the rules of determinant manipulations apply to these 
equations, and we may evaluate many problems analytically.

\vfill\eject
Let us summarize the main conclusions of this section:

\bigskip
$\bullet$ The \underbar{Eulerian evolution equations} for the velocity and 
density fields (1a) and (1b) are integrated exactly in the Lagrangian picture
by (2d) and (2e), respectively.

\bigskip
$\bullet$ The transformation of the 
\underbar{Eulerian field equations} (1c,d) yields
a system of \underbar{Lagrangian field equations} 
for the acceleration field.

\bigskip
$\bullet$ Using the integrals for the acceleration and density fields
(2d) and (2e)
we arrive at the \underbar{Lagrange--Newton system} (8) 
which is a closed system of \underbar{Lagrangian evolution equations} for the 
\underbar{field of trajectories}. 
Density and velocity are no longer dynamical variables,
but are replaced by the \underbar{single} dynamical field $\vec f (\vec X,t)$.

\bigskip
$\bullet$ The Lagrange--Newton system (8) 
is \underbar{equivalent} to the 
Euler--Newton system (1) as long as the mapping $\vec f$ is 
\underbar{invertible} ($J>0$) and, in particular, 
\underbar{non--singular} ($J \ne 0$); 
(for a proof see $\lbrack 21 \rbrack$).   
Note that the Lagrange--Newton system remains regular at caustics ($J=0$) where 
the Eulerian representation breaks down. 
Whether the Lagrangian equations still describe the physics correctly in the 
regime $J<0$ will be discussed in Section 3.

\bigskip

\section{Lagrangian Dynamics of Eulerian Fields}

The question how to obtain the evolution of Eulerian fields from the 
Lagrangian description is easily answered: Knowing the 
Lagrangian solution of the transformation (2a), $\vec x = \vec f (\vec X,t)$,
we have to express the Eulerian fields first in terms of $\vec f$, and then
use the inverse of this transformation (2g), $\vec X = \vec h (\vec x,t)$,
to map the variable $\vec X$ back to Eulerian space, e.g., for the velocity
we have:
$$
\vec v (\vec x,t) = \dot{\vec f}(\vec h(\vec x,t),t)\;\;.\eqno(9)
$$
\smallskip
\noindent
The power of a Lagrangian description mainly relies on this implicit 
determination of the evolution of Eulerian fields. 
However, we can only come back to Eulerian space as long as
$\vec h$ exists. In general, the inverse transformation can be multivalued
(see the discussion in Section 3).

The same rule applies to {\it any} Eulerian field, so we only have to find
the corresponding formula which expresses it in terms of $\vec f$.

\bigskip\noindent
Let us now discuss another way of describing the evolution of Eulerian fields
along trajectories. This will lead us to equations 
which are frequently discussed in the literature.
These equations are
useful to illuminate the power of the Lagrangian formalism outlined in Section 
1, but they do not determine the evolution of Eulerian fields {\it per se},
as will be shown below.

We may ask for evolution equations which involve the Lagrangian time--derivative
(2c) instead of the Eulerian time--derivative $\partial_t$ as in (1a) and (1b). 
They may be written in symbolic form as:
$$
\dot{\cal E}_{\nu}= {\cal F}_{\nu}({\cal E}_{\mu})\;\;,
$$
i.e., a system of $\nu$ equations which determines the evolution of the $\nu$ variables
${\cal E}_{\nu}$
along the flow lines. A solution of this system of equations does not
provide the full answer, which needs knowledge of the trajectories 
themselves! We will see that the procedure outlined at the beginning of this
subsection {\it does provide} the full answer. We shall derive now
{\it Lagrangian evolution equations} for a number of fields of interest, 
and shall then reinforce the Lagrange--Newton system (8). 

We start with eq. (1a) and write it down in index notation,
$$
\partial_t \; v_i + v_k v_{i,k} = g_i \;\;.\eqno(10a)
$$
Performing the spatial Eulerian derivative of this equation and using (2c) we obtain:
$$
(v_{i,j})^{\bdot} = - v_{i,k}v_{k,j} + g_{i,j}\;\;.
\eqno(10b)
$$
Eq. (10b) is an evolution equation for the velocity gradient $(v_{i,j})$
along the flow lines $\vec f$. 
It is convenient for the discussion of fluid motions to split it into 
its symmetric part (expansion tensor $\theta_{ij}$), 
its antisymmetric part (vorticity tensor $\omega_{ij}$), 
and to separate the symmetric part into a tracefree
part (the shear tensor $\sigma_{ij}$) and the trace 
(the rate of expansion) $\theta: = v_{i,i}$:
$$
v_{i,j} = v_{(i,j)} + v_{[i,j]}=:\theta_{ij} + \omega_{ij} 
=:\sigma_{ij} + {1\over 3}\theta\delta_{ij} + \omega_{ij}
\;\;,\eqno(11)
$$
where $v_{(i,j)}={1\over 2}(v_{i,j}+v_{j,i})$ and $v_{[i,j]}={1\over 2}(v_{i,j}-v_{j,i})$.

\noindent
Inserting (11) into (10b) we obtain the following evolution equations
(compare $\lbrack 20 \rbrack$, $\lbrack 22 \rbrack$, $\lbrack 32 \rbrack$, 
$\lbrack 28 \rbrack$(\S 22):
$$
\eqalignno{
&\dot\theta = -{1\over 3}\theta^2 + 2(\omega^2 -\sigma^2 ) + g_{i,i}\;\;,
&(12a)\cr
&(\omega_{ij})^{\bdot}=
-{2\over 3}\theta \omega_{ij} -\sigma_{ik}\omega_{kj}
-\omega_{ik}\sigma_{kj}+g_{[i,j]}\;\;,
&(12b)\cr
&(\sigma_{ij})^{\bdot}=
-{2\over 3}\theta \sigma_{ij} -\sigma_{ik}\sigma_{kj}
-\omega_{ik}\omega_{kj}+{2\over 3}(\sigma^2 - \omega^2 )\delta_{ij}
+g_{(i,j)}-{1\over 3}g_{k,k}\delta_{ij}\;\;,
&(12c)\cr}
$$
where $\sigma^2 := {1\over 2}\sigma_{ij}\sigma_{ij}$ and $\omega^2 := {1\over 2}
\omega_{ij}\omega_{ij}$.

We can read these equations in the sense that they reconstruct 
the acceleration gradient $(g_{i,j})$ in terms of kinematical variables: 
eq. (12a) gives the 
trace of $(g_{i,j})$, eq. (12b) its antisymmetric part, and eq. (12c) its 
tracefree symmetric part, the {\it Newtonian tidal tensor} $E_{ij}:=g_{i,j}-
{1\over 3}g_{k,k}\delta_{ij}$.

We may now establish a system of equations for the variables $\varrho$, 
$\theta$, $\sigma_{ij}$ and $\omega_{ij}$ by using the Euler--Newton system
(1). This system only constrains the trace and the antisymmetric part of 
($g_{i,j}$), but not the tidal tensor $E_{ij}$; we arrive at:
$$
\eqalignno{
&\dot\varrho= -\varrho\theta\;\;,&(13a)\cr
&\dot\theta = -{1\over 3}\theta^2 + 2(\omega^2 -\sigma^2 ) + \Lambda - 4\pi G
\varrho\;\;,&(13b)\cr
&(\omega_{i})^{\bdot}=-{2\over 3}\theta \omega_i +\sigma_{ij}\omega_j
\;\;,&(13c)\cr
&(\sigma_{ij})^{\bdot}=
-{2\over 3}\theta \sigma_{ij} -\sigma_{ik}\sigma_{kj}
-\omega_{ik}\omega_{kj}+{2\over 3}(\sigma^2 - \omega^2 )\delta_{ij}
+ E_{ij}\;\;;&(13d)\cr}
$$
(we have expressed the vorticity tensor in terms of the vector
$\vec \omega = {1\over 2}\nabla \times \vec v$ by means of the formula
$\omega_{ij} = -\epsilon_{ijk}\omega_k$).

There have been efforts in the literature to close this system of 
{\it ordinary} differential equations by using the corresponding equations
of General Relativity (compare $\lbrack 1 \rbrack$ and 
the lecture by Matarrese) and looking at their
Newtonian limits. (In fact, the eqs. (13a--d) are formally identical to their GR
counterparts in ``comoving'', i.e., Lagrangian coordinates.)
However, it turns out that, although we can formally 
derive an evolution equation for the tidal tensor, the system of equations
(13) supplemented by the evolution equation for the tidal tensor
is not a system of {\it ordinary} differential equations and the problem 
remains ``non--local'' $\lbrack 2 \rbrack$, $\lbrack 24 \rbrack$, 
$\lbrack 23 \rbrack$, $\lbrack 21 \rbrack$.

As a matter of fact, in Newtonian theory we don't need evolution equations for 
tracefree symmetric tensors (like the shear and tidal tensors) 
to get a closed system of equations.
We have already obtained such a system without them: the Lagrange--Newton
system (8), which is a set of {\it partial} differential equations.

\noindent  
To see the relation of the system of equations (13) to the Lagrange--Newton 
system we can show the following (the proof I leave to the reader 
as an excercise):

\medskip\noindent
$\imath$.) Eq. (13a), the {\it continuity equation}, 
is equivalent to eq. (1b) and is integrated in the Lagrangian
framework by (2e).

\smallskip\noindent
$\imath\imath$.) Eq. (13b), 
{\it Raychaudhuri's equation}, is equivalent to eq. (1d),
if $\vec g$ is related to the velocity as in eq. (1a).

\smallskip\noindent
$\imath\imath\imath$.) Eq. (13c), 
the {\it Kelvin--Helmholtz vorticity transport equation}, 
is equivalent to eq. (1c), 
if $\vec g$ is related to the velocity as in eq. (1a).

\medskip
Note that eq. (13c) can also be integrated exactly in the Lagrangian picture, 
a result due to Cauchy (see $\lbrack 31 \rbrack$  -- a good textbook on Lagrangian dynamics):
$$
\vec{\omega} = {1\over J}\vec{\omega}(\vec X,t_0)\cdot\nabla_0 \vec f \;\;.\eqno(14)
$$

\bigskip\medskip

We arrive at the following conclusions of this section:
\bigskip

$\bullet$
The equations (13a--c) are \underbar{equivalent} 
to the equations (1b--d) provided the relation 
between velocity and acceleration
is given by (1a).

\bigskip
$\bullet$
The equations (13b,c), if expressed in terms of Lagrangian coordinates,
yield a \underbar{closed Lagrangian system}, the
Lagrange--Newton system (8a--d).

\bigskip
$\bullet$
The evolution of the tracefree symmetric tensors $\sigma_{ij}$ and 
$E_{ij}$ can be calculated {\it after} a solution  
to the Lagrange--Newton
system is obtained, the formulas can be read off from eq. (6) 
(and a similar equation for the velocity gradient): 
$$
\eqalignno{
&\sigma_{ij} = {1\over 2J} \epsilon_{jpq}{\cal J}({\dot f}_i,f_p,f_q)
- {1\over 6 J}\epsilon_{opq}{\cal J}({\dot f}_o,f_p,f_q)\delta_{ij}
\;\;;&(15a)\cr
&E_{ij} = {1\over 2J} \epsilon_{jpq}{\cal J}({\ddot f}_i,f_p,f_q)
- {1\over 6 J}\epsilon_{opq}{\cal J}({\ddot f}_o,f_p,f_q)\delta_{ij}\;\;&(15b)\cr
&\;\;\;\;\;\;={1\over 2J} \epsilon_{jpq}{\cal J}({\ddot f}_i,f_p,f_q)
-{1\over 3}\left(\Lambda - {4\pi G \over J}
\ueber{\varrho}{o}\right)\delta_{ij}
\;\;.&(15c)\cr}
$$
We could use eq. (15c) to give another way of stating the Lagrange--Newton 
system: it is equivalent to the conditions that $E_{ij}$ is symmetric and  tracefree
(compare (7)):
$$
\eqalign{
&E_{[i,j]} = 0 \;\;\Leftrightarrow\;\;(8a,b,c)\;\;;\cr
&E_{ii} = 0 \;\;\Leftrightarrow\;\;(8d)\;\;.\cr}
$$
\bigskip

$\bullet$
The formulas (14) and (15) are examples of formulas which express some 
Eulerian field 
quantity in terms of $\vec f$. Given any solution for $\vec f$, we can insert it and
use the inverse of the same solution to map the field into Eulerian space 
as in (9). (Integrals of the Eulerian evolution equations along the trajectories
without knowledge of the trajectories themselves do not provide this information
directly.)
The perturbation solutions discussed in the next section will provide examples.

\bigskip

\section{Lagrangian Perturbation Theory}

The previous sections have equipped us with the necessary framework
in which the Lagrangian dynamics of self--gravitating flows 
can be studied. We have reduced the description of the dynamics of any
Eulerian field
to the problem of finding the field of trajectories $\vec f$ as a solution of the 
Lagrange--Newton system (8).
This problem will be addressed now.

\medskip

\subsection{ The Lagrangian perturbation approach}

As in the Eulerian case we are not able to write down a solution for 
generic initial data (i.e. without any symmetry assumptions like
plane or spherical symmetry). 
We may start with the simplest class
of solutions, the homogeneous--isotropic ones, and then investigate a perturbative 
treatment of inhomogeneities. For homogeneous solutions 
the deformation tensor $(f_{i\1 k})$ does not depend on 
$\vec X$ and the trajectory field is that of a {\it homogeneous
deformation}:
$$
f^H_i (\vec X,t) = a_{ij}(t) X_j \;\;;\;\;a_{ij} (t_0):=\delta_{ij} \;\;;\eqno(16a,b)
$$
we assume hereafter that it is isotropic,
$$
a_{ij}(t)=a(t)\delta_{ij}\;\;.\eqno(16c)
$$
Inserting this ansatz into the Lagrange--Newton system (8) yields the 
well--known equation
$$
{\ddot a \over a} = {\Lambda - 4\pi G \varrho_H \over 3}\;\;,\eqno(16d)
$$
where $\varrho_H = \varrho_H (t_0)a^{-3}$ is the homogeneous density field
calculated from (16a,c) and (2e).
We may use it to integrate eq. (16d) yielding Friedmann's differential 
equation:
$$
{{\dot a}^2 - e \over a^2} =
{8 \pi G \varrho_H + \Lambda \over 3} \; ; \; e =const
\;\;.\eqno(16e)
$$
Solutions of (16e) cover the standard Friedmann cosmologies.
We now define the (not necessarily small) deviation $\vec p$ from the 
``background model'' ${\vec f}^H$, so that the full displacement map
$\vec f$ of an inhomogeneous model reads:
$$
\vec f = \vec f^H + \vec p (\vec X,t)\;\;;\;\;\vec p (\vec X,t_0):=\vec
0  \;\;.\eqno(17a,b)
$$
(Eq. (17b) expresses the freedom of starting at the same initial position as in the 
homogeneous case.)

\noindent
The {\it Lagrangian perturbation approach} then consists of solving the 
Lagrange--Newton system for powers of $\varepsilon$ 
to obtain the $m$th order solution:
$$
{\vec f}^{\,\lbrack m \rbrack} = \vec f^H + {\vec p}^{\,\lbrack m \rbrack}
\;\;\;\;;\;\;\;\;
{\vec p}^{\,\lbrack m \rbrack} = \sum_{i=1}^m \;\varepsilon^i
{\vec p}^{\,(i)}\;\;.\eqno(17c,d)
$$
Note the following important differences to the Eulerian perturbation approach:
1. we need not perturb the density field as in the Eulerian case; 
2. the perturbed flow $\vec f$ is expressed in terms of coordinates which are constant 
along this flow, while in the Eulerian case the perturbation is expressed in a fixed
frame, which does not change by moving away from the homogeneous flow. 
This implies that, although deviations $\vec p$ from ${\vec f}^H$ might be small
compared to the homogeneous displacement,  
the Eulerian fields evaluated along the perturbed flow can experience large changes
(Fig.2). 

\begin{figure} 

\psfig{figure=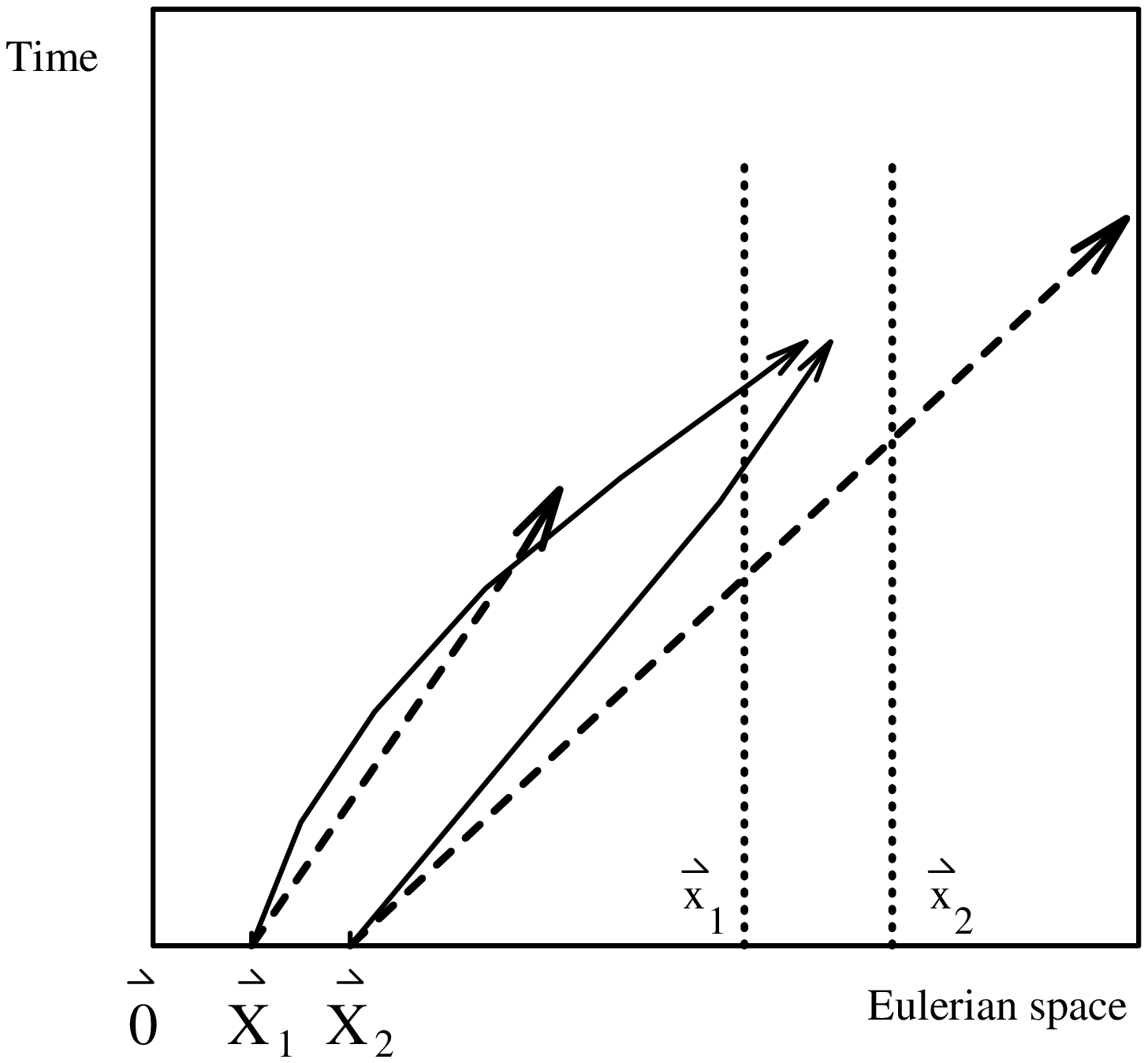,height=7.0cm,width=7.0cm}
\vskip -7.1 true cm
\rightline{\psfig{figure=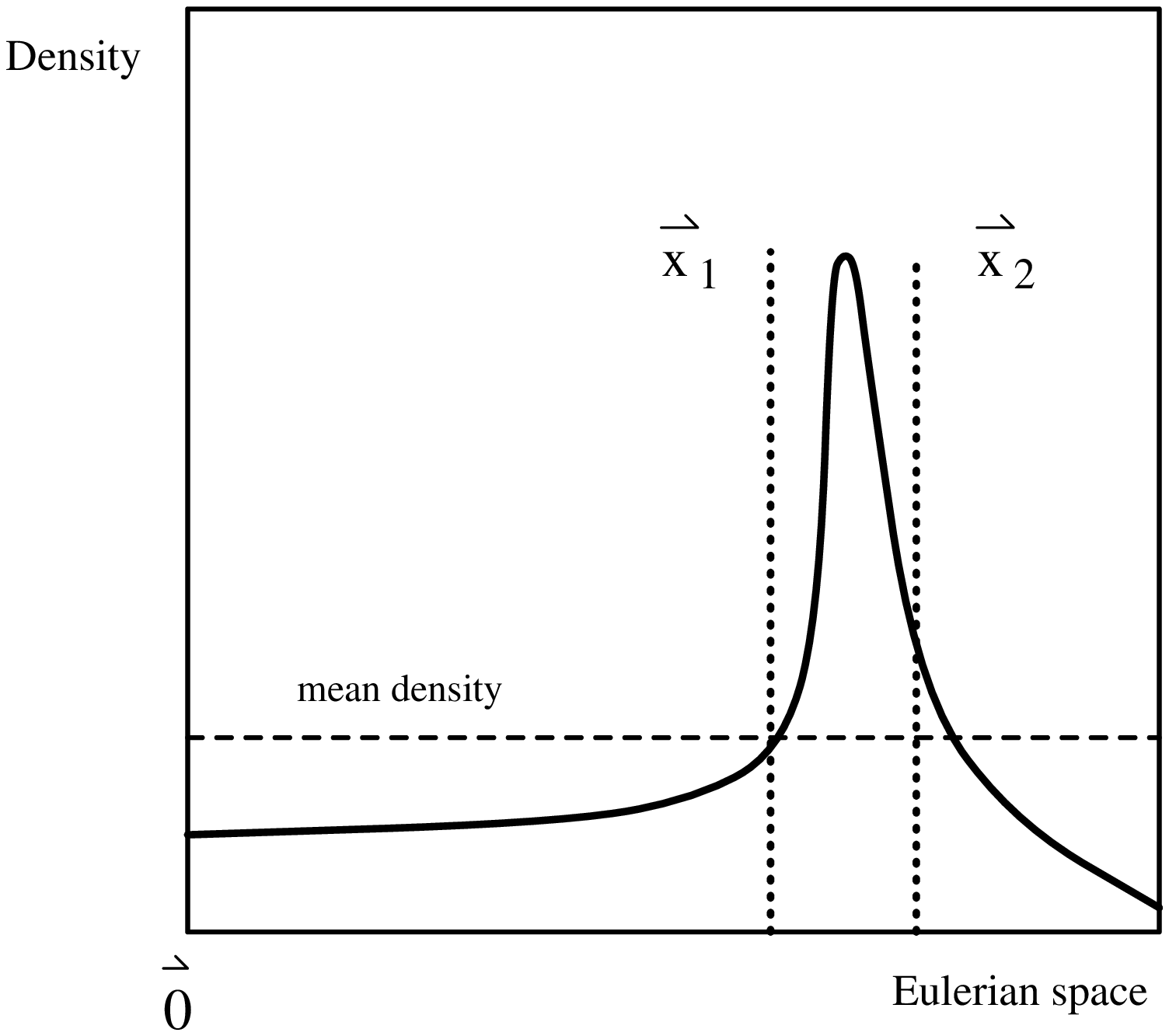,height=7.0cm,width=7.0cm}}

\caption{
The unperturbed Hubble--flow is represented by the motion of two fluid particles
(dashed lines) and the perturbed flow is superimposed (full lines), Fig.2a. In spite of 
the smallness of the deviations from the Hubble--flow compared to the Hubble--displacements, 
the density in Eulerian space
can experience large changes, Fig.2b.}

\end{figure}
 
Solutions of the form (17) can be found and in the following subsections we shall learn
more about them. 

\vskip -1 true cm
\subsection{A comment on average flows}

In the ansatz (17) we have splitted the inhomogeneous model into a homogeneous--isotropic
model and a deviation from it.
Before we discuss solutions of the form (17) we have to look at another 
assumption which has to be imposed on the inhomogeneous models. This will be 
necessary for their application to cosmology.

In cosmology we want   
the homogeneous--isotropic deformation to describe the {\it mean motion} of the whole
universe, and we want to interprete this motion as a reference frame (associated
with, e.g., the microwave background).  
This interpretation requires that the background model (which, so far, is a homogeneous solution)
correctly describes the
{\it average flow}, i.e., inhomogeneities or peculiar--motions must vanish on average 
on some large scale. This we call the {\it cosmological principle of homogeneity}.
That the average flow is isotropic is another assumption which we set {\it a priori}.
It is, however, supported by the extreme isotropy of the microwave background.
Therefore, in order to make Lagrangian perturbation models applicable to reality, we have to
carefully analyze the average of inhomogeneous models. For this purpose we may look at 
the general expansion law in Newtonian cosmology by averaging  
the system of equations (13). A general discussion is to be found in  
$\lbrack 15 \rbrack$. 
This study shows that the general expansion law for an expansion function 
$a_{\cal D}:=V^{1/3}$, where $V$ is the volume of the spatial domain ${\cal D}(t)$ 
on which the average
is performed, is of the form:
$$
3{{\ddot a}_{\cal D} \over a_{\cal D}} + 4\pi G {M_{\cal D}\over a_{\cal D}^3} -
\Lambda = 
-{2\over 3}\left(a^{-3}_{\cal D}\int_{{\cal D}(t)}\nabla\cdot\vec u \right)^2
+ a^{-3}_{\cal D}\int_{{\cal D}(t)}\nabla\cdot{\vec {\cal Q}}\;\;,\;\;
{\vec {\cal Q}}: = \left(\vec u \,\nabla\cdot \vec u - \vec u \cdot
\nabla \vec u \right)\;\;,\eqno(18a)
$$
where $M_{\cal D}$ is the total mass contained in ${\cal D}(t)$. 
Eq. (18a) shows that the average expansion is of Friedmann type, 
$a_{\cal D}\equiv a$, only in the case where
the averages of the nonlinear terms on the r.h.s. (which depend on the 
inhomogeneities) vanish (compare (16d)).
In other words, if we follow inhomogeneities into the nonlinear regime (and that's what
we have in mind with Lagrangian approximations), then 
we actually neglect the influence of the 
inhomogeneities on the global expansion when we use homogeneous solutions for the 
average flow. Since these nonlinear source terms can be written as a divergence
$\lbrack 15 \rbrack$, we can use Gau{\ss}'s theorem to transform 
the volume integrals
in (18a) into surface integrals over the boundary of the domain 
$\partial{\cal D}$, e.g.:
$$
\int_{{\cal D}(t)} d^3 x \;\nabla \cdot {\vec {\cal Q}} = 
\int_{\partial{\cal D}(t)}
\vec{dS}\cdot{\vec {\cal Q}}\;\;.\eqno(18b)
$$ 
The flux of $\vec u$ and $\vec{\cal Q}$ through 
the boundary of the averaging volume is assumed to be negligible. 
This can only be true, if the extent of structures 
in the Universe is considerably smaller than the averaging domain. 
In standard cosmologies we assume that this is indeed the case on some large scale
where we apply our models, 
and we even assume that
this boundary flux is exactly zero.

It is interesting to see at this stage that we implicitly satisfy this assumption 
by imposing
{\it periodic boundary conditions} on the cosmic peculiar--fields: 
Periodicity means that we can
alternatively view the Universe as a torus. A torus is a compact space without 
boundary, and consequently any flux through the boundary vanishes exactly.

\medskip

\subsection{Perturbation solutions at first order}

In this subsection we shall write down the {\it general} 
first--order solution in order to see
that we need quite a list of arguments to reduce it to what we call the
``Zel'dovich--approximation'' $\lbrack 34 \rbrack$, $\lbrack 35 \rbrack$.

A solution to the basic system of equations (8) is {\it general},
if we can specify initial data for the velocity
{\it independently} of a given
density field (or acceleration field, respectively). 
If we split these data into their homogeneous
parts (belonging to the background solution) and deviations
from the background, {\it peculiar--velocity} $\vec u$ and {\it peculiar--acceleration}
$\vec w$, then we have to specify
$$
\vec U := \vec u (\vec X, t_0) =:{\vec U}^L + {\vec U}^T \;\;;\;\;
\vec W := \vec w (\vec X, t_0)\;\;,\eqno(19a)
$$ 
where we have splitted the peculiar--velocity into its irrotational (longitudinal) part
${\vec U}^L$ and its rotational (transverse) part ${\vec U}^T$ (the acceleration is
irrotational in Newton's theory). We may introduce scalar potentials and a vector potential
such that
$$
{\vec U}^L =:\nabla_0 S\;\;;\;\;{\vec U}^T = :\nabla_0 \times {\vec S}\;\;;\;\;
\vec W =:\nabla_0 \Phi \;\;.\eqno(19b)
$$
Solving the Lagrange--Newton system to the first order with the perturbation ansatz 
(17) we obtain $\lbrack 9 \rbrack$:
$$
{\vec f}^{\lbrack 1\rbrack}=a(t)\vec X + b_1 (t) \nabla_0 S^{(1)}(\vec X)
               + b_2 (t) \nabla_0 \times {\vec S}^{(1)}(\vec X)  
               + b_3 (t) \nabla_0 {\Phi}^{(1)}(\vec X)\;\;,\eqno(20)
$$
where the $b_i (t)$ depend on the chosen background solution $a(t)$, 
$b_i (t_0)=0$, and 
the $\vec X$--dependent perturbation potentials have to obey Poisson equations which 
relate them to the initial data (19b):
$$
\Delta_0 S^{(1)} = \Delta_0 S t_0 \;\;;\;\; 
\Delta_0 {\vec S}^{(1)} = \Delta_0 {\vec S} t_0 \;\;;\;\;\Delta_0 {\Phi}^{(1)} 
= \Delta_0 {\Phi} t_0^2\;\;.\eqno(20a-e) 
$$
($\Delta_0$ denotes the Laplace operator with respect to Lagrangian coordinates.)

\noindent
The solution (20) shows that the evolution of a fluid element depends (non--locally) on 
the distribution of all fluid elements, because Poisson equations have to be solved
(which is clear in view of the structure of the 
theory). The rotational part stems from the Lagrangian equations (8a--c) which must not
be forgotten! Only at first order the rotational part directly represents 
rotational flow in Eulerian space. At higher orders rotational parts arise 
also for irrotational flow.
This can be explained by the following consideration: Imagine an arrow for which
we require that it keeps a fixed angle to some axis in an 
Eulerian coordinate system. 
The arrow
may be placed on a system which is orbiting around some point. In this 
(Lagrangian) system the arrow performs a rotating movement, while in Eulerian
space it is just translationally shifted. Indeed, if 
we restrict the motion to be irrotational in Lagrangian space, then this 
restricts possible classes of motion far more than the requirement of irrotational 
motion in Eulerian space.   

Let us return to the solution (20). 
At first order it is straightforward to solve the $5$ Poisson equations (20a--e) 
explicitly:
$$
S^{(1)}= S t_0 + \Psi \;\;;\;\;
{\vec S}^{(1)}={\vec S}t_0 + {\vec \Psi}\;\;;\;\;
{\Phi}^{(1)} =\Phi t_0^2+ \Omega\;\;,
$$
$$
\Delta_0 \Psi = 0\;\;;\;\;\Delta_0 {\vec \Psi}=\vec 0\;\;;\;\;
\Delta_0 \Omega =0\;\;.\eqno(21a-e)
$$
We can get rid of the harmonic functions $\Psi$, $\vec\Psi$ and $\Omega$
and we can make $S$, ${\vec S}$ and $\Phi$ unique, if 
we impose periodic boundary conditions on the initial data: the only periodic harmonic
functions are constant in space, a constant which expresses a translation of the whole
simulation box and which can be set to zero, since the basic equations are translationally
invariant (see $\lbrack 12 \rbrack$, Appendix C and $\lbrack 21 \rbrack$).
If we don't do that then the solutions are not unique!
Removing this freedom reduces (20) to a {\it local} approximation which is 
only possible at first order.

For the purpose of simulating the evolution of large--scale structure forward in time
we can simplify the solution further by assuming that, initially
$$
\vec U (\vec X) = \vec W (\vec X) t_0 \;\;,\eqno(22)
$$
i.e., peculiar--velocity and --acceleration 
are parallel to start with (in particular, we have 
${\vec U}^T (\vec X) = \vec 0$). This assumption is justified
on physical grounds: as long as the perturbations are small, we may describe them
by the (Eulerian) linear theory. Then, asymptotically (after the decaying modes have died
out) the condition (22) is approximately 
reached and we may start with this condition from the outset.
Employing (22) the first--order solution reduces to the ``Zel'dovich--approximation''
(compare $\lbrack 9 \rbrack$ for the Einstein--de Sitter background, and 
$\lbrack 8 \rbrack$, $\lbrack 4 \rbrack$
for general backgrounds including a cosmological constant 
to see that the time--coefficients come out right):
$$
{\vec f}^{\lbrack Z\rbrack}=a(t)\vec X + b (t) \nabla_0 S (\vec X) \;\;.\eqno(23)
$$
Note that, strictly speaking, (23) is not the growing part of the general solution, but is that part
which results from the restriction (22) on initial data: this restriction automatically
cancels the decaying modes at first order, but not at higher orders.

\medskip

\subsection{Higher--order solutions}

The solution of the Lagrange--Newton system for higher orders in $\varepsilon$ for the initial data
(22) systematically gives us higher--order corrections to the 
``Zel'dovich--approximation''. For {\it general} initial data, already the second--order
solution needs four pages to show it (compare $\lbrack 11 \rbrack$ for the 
longitudinal part, which comprises the solution of one of the four equations 
(8a--d)). 
Restricting them to the setting (22) we obtain the much
simpler form which is commonly being
studied ($\lbrack 5 \rbrack$, compare Bouchet's
lecture, and $\lbrack 10 \rbrack$, $\lbrack 11 \rbrack$ 
for alternative models of this form):
$$
{\vec f}^{\lbrack 2\rbrack}=a(t)\vec X\;+\;
q_1 (t) \; \nabla_0 {\cal S}^{(1)} (\vec X) \;+\; q_{2}(t) 
\; \nabla_0 {\cal S}^{(2)} (\vec X)
\;\;.\eqno(24)
$$
The perturbation potentials in (24) have to be constructed by
solving iteratively the two Poisson equations:
$$
\eqalignno{
&\Delta_0 {S}^{(1)} = I({S}_{\1 i\1 k})t_0  \;\;\;,&(24a)
\cr
&\Delta_0 {S}^{(2)} = 2 II({S}^{(1)}_{\1 i\1 k})
\;\;\;. &(24b) \cr}
$$
$I$ and $II$ denote the first and second principal
scalar invariants of the tensor gradient in brackets:
$$
I({S}_{\1 i\1 k}) = tr({S}_{\1 i\1 k}) =
\Delta_0 {S} \;\;\;, \eqno(24c)
$$
$$
II({S}^{(1)}_{\1 i\1 k}) = {1 \over 2}
\lbrack(tr({S}^{(1)}_{\1 i\1 k}))^2 -
tr(({S}^{(1)}_{\1 i\1 k})^2)\rbrack \;\;\;.\eqno(24d)
$$
The third--order solution for the setting (22) may be found in 
$\lbrack 12 \rbrack$,
the time--coefficients being evaluated for an Einstein--de Sitter background.
(For other backgrounds we cannot write them in explicit form, see 
$\lbrack 18 \rbrack$ for a thorough discussion.) Illustrations are given 
in Figs. 3,4.

\begin{figure}

\psfig{figure=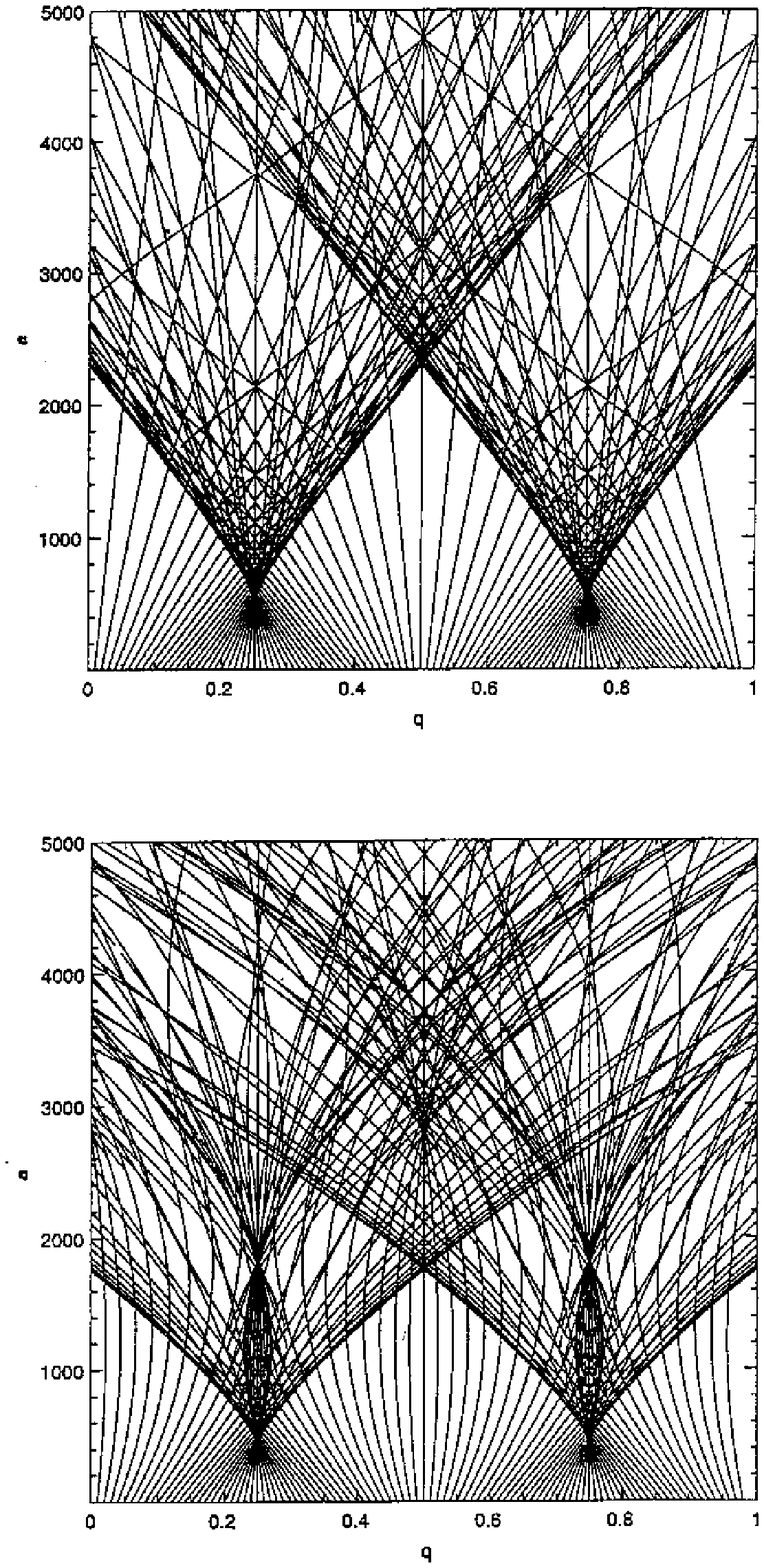,height=12.0cm,width=6.0cm}
\vskip -12.2 true cm
\rightline{\psfig{figure=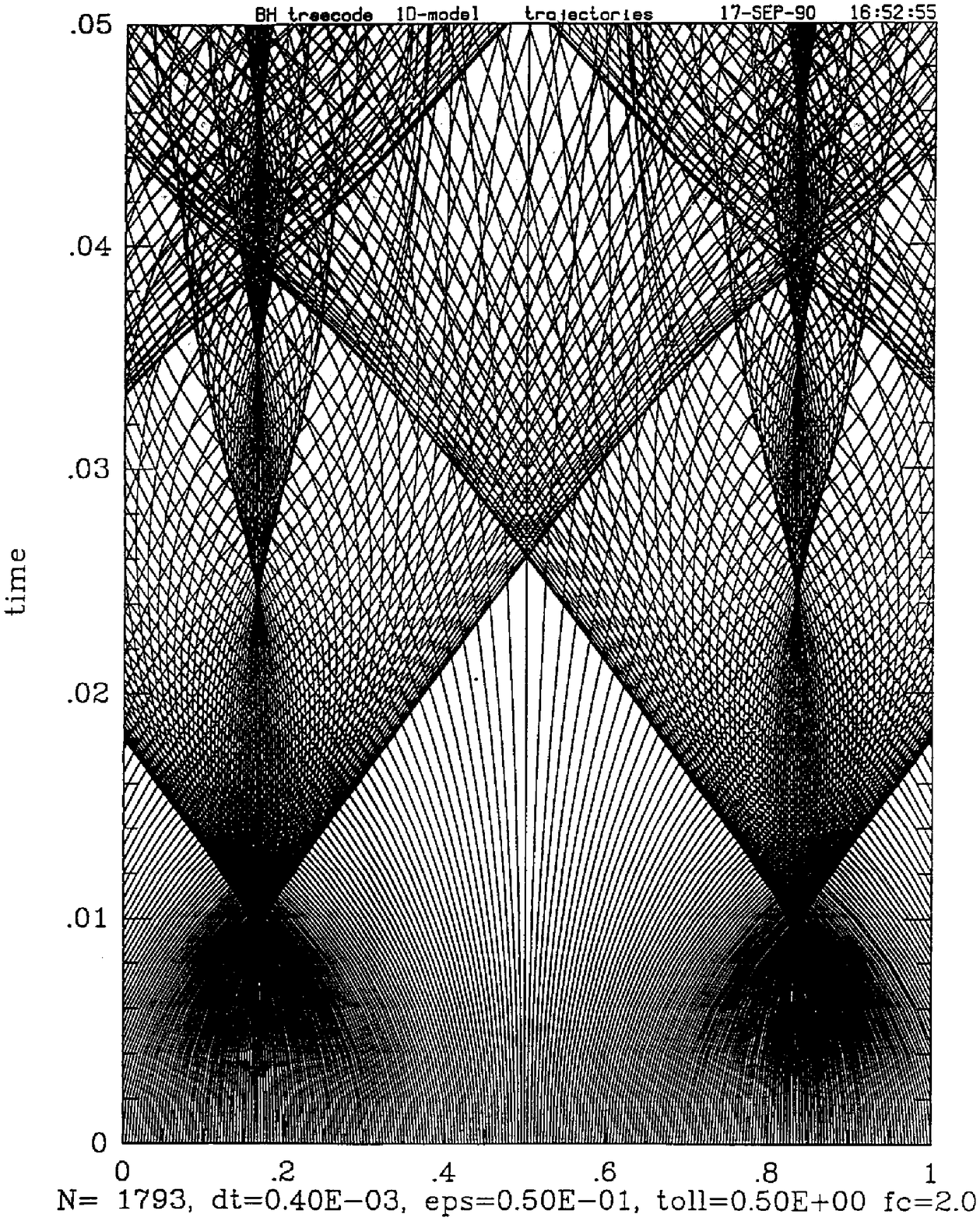,height=12.0cm,width=8.0cm}}

\caption{
The trajectories for a simple plane wave model (2D) 
are plotted for the first--order 
(left, top panel) and for the second--order solution (left, bottom panel) 
(taken from $\lbrack 11 \rbrack$) compared with the trajectories in 
a two--dimensional tree--code simulation (taken from $\lbrack 17 \rbrack$).
Those trajectories are plotted which end in an Eulerian spacetime section to show that
the second--order model displays a secondary shell--crossing (this is not visible
for trajectories which lie in this plane for all times -- they follow a 
plane--symmetric collapse 
for which the first--order model is the exact solution). Two generations of 
caustics successively appear, whereas the numerical simulation shows 
further generations.}
 
\end{figure}

\begin{figure}

\psfig{figure=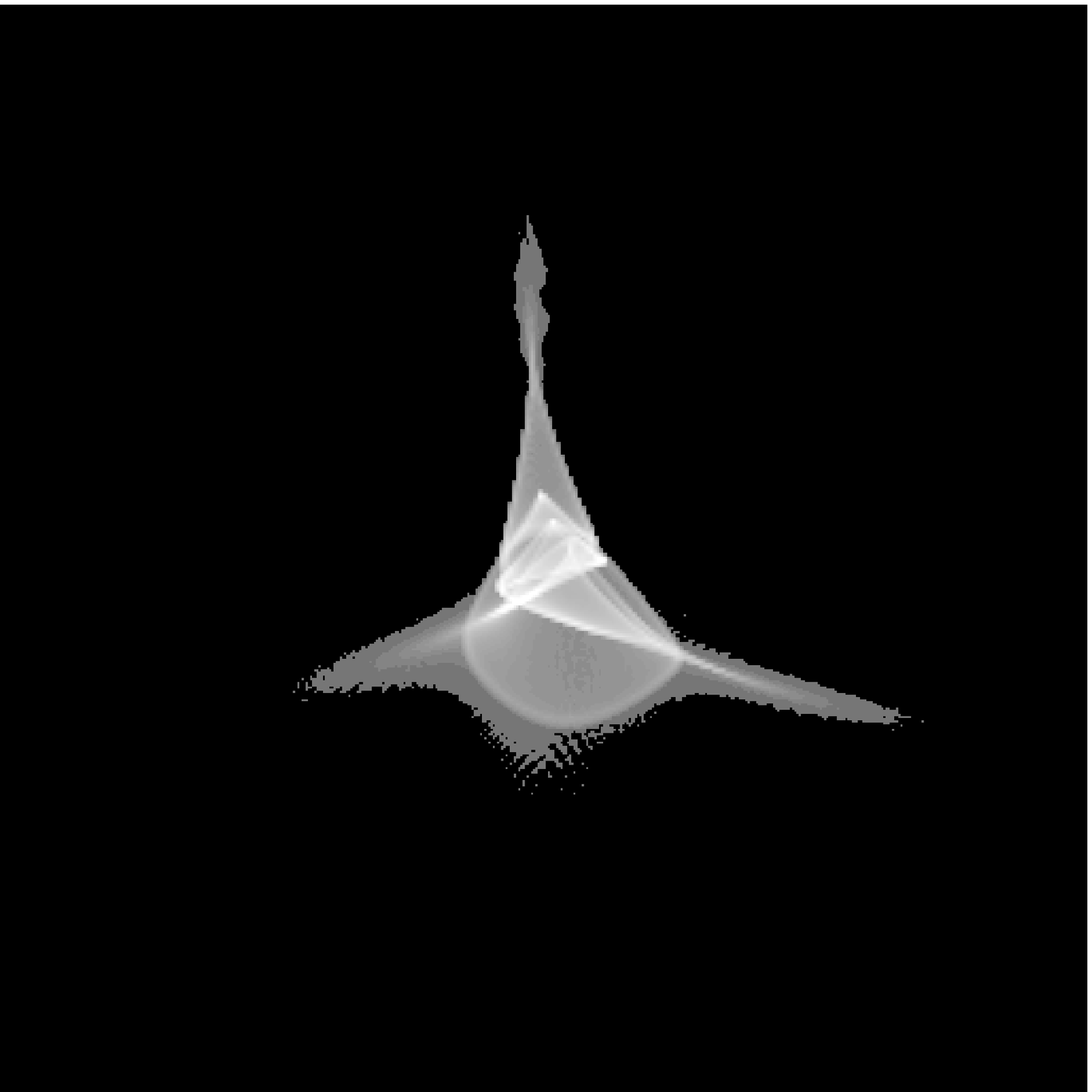,height=7.0cm,width=7.0cm}
\vskip -7.1 true cm
\rightline{\psfig{figure=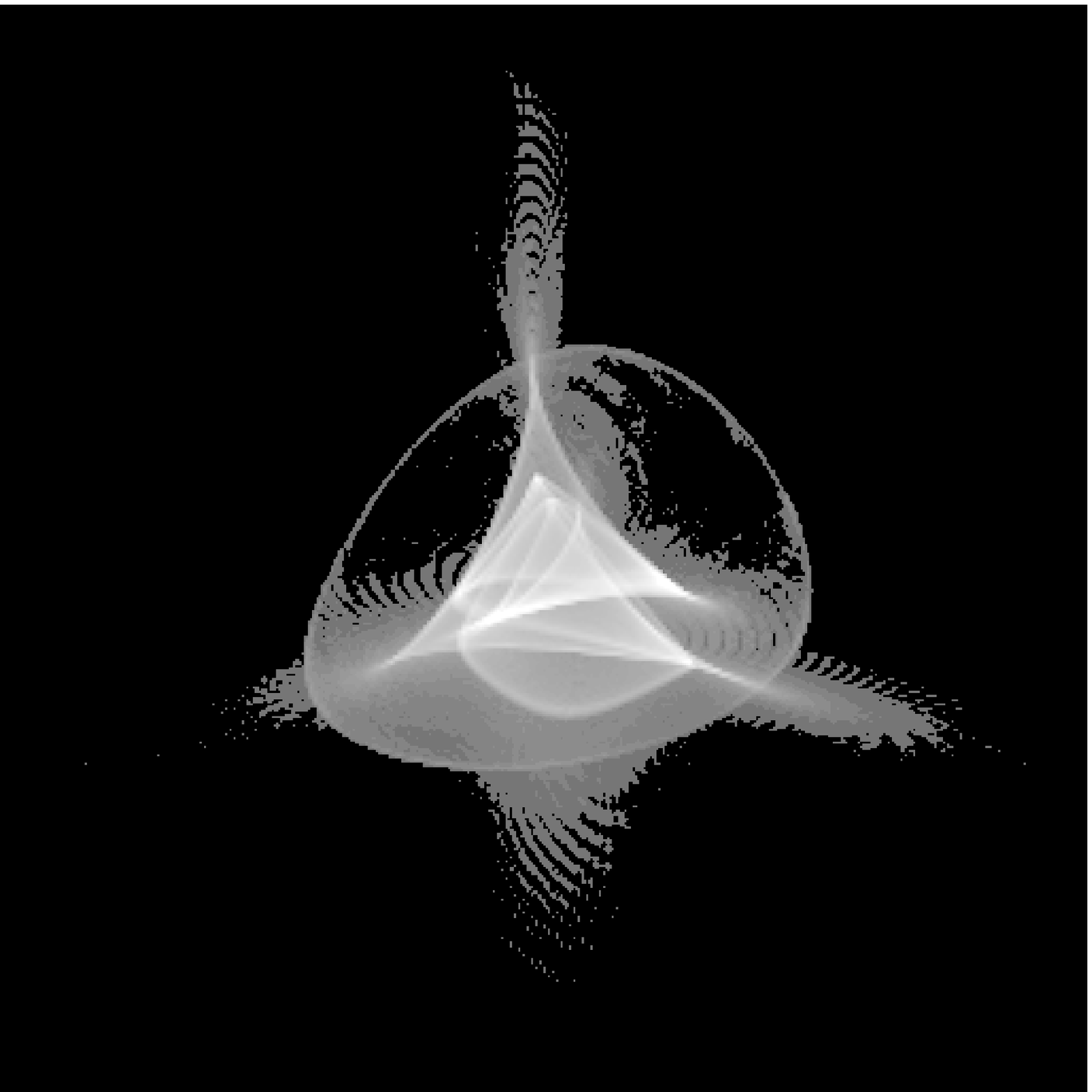,height=7.0cm,width=7.0cm}}

\caption{A slice through the density fields ($512^3$ trajectories collected into a $256^3$ pixel 
grid) predicted by the Lagrangian approximations at first order (left)
and third order (right panel) for a 3D plane wave model is shown 
(taken from $\lbrack 17 \rbrack$).
The second-- and third--order schemes feature secondary shell--crossings. 
This hierarchy of nested caustics continues down to smaller scales as was shown in
a two--dimensional numerical simulation in $\lbrack 19 \rbrack$.}

\end{figure}

\subsection{Comparison of Eulerian and Lagrangian perturbation theory}

For the comparison of both perturbation approaches we look at
the equations governing the evolution of inhomogeneities
in the first--order Eulerian and Lagrangian approximations.
Consider the
contrast density $\Delta: = (\varrho - \varrho_H) / \varrho$,
$- \infty < \Delta < 1$,
which is more adapted to the nonlinear situation than the conventional
definition $\delta=(\varrho - \varrho_H) / \varrho_H = \Delta /
(1 - \Delta)$, defined in Eulerian perturbation theory.
For this field we can find the following exact
evolution equations by inserting the continuity equation into Raychaudhuri's
equation (13b) and splitting off the homogeneous part 
$\lbrack 29 \rbrack$, $\lbrack 8 \rbrack$, $\lbrack 9 \rbrack$:
$$
\eqalignno{
&\dot \Delta = (\Delta-1) I \;\;, &(25a) \cr
&\ddot \Delta + 2 {\dot a \over a} \dot \Delta -
4 \pi G \varrho_H \Delta =
(\Delta-1) 2 II \;\;,  &(25b) \cr}
$$
where $I$ and $II$
denote here the invariants
of the peculiar--velocity tensor gradient with respect to
Eulerian coordinates which are scaled by the expansion factor, 
$q_i = x_i / a(t)$: $(\partial u_i / \partial
q_j)$.

For $II = 0$ the equations (25) are
(except for the Lagrangian time derivative and the redefinition of
the density contrast) the same equations as known in Eulerian linear
theory for
$\delta (\vec q,t)$ $\lbrack 28 \rbrack$:
$$
\eqalignno{
&{\partial \over \partial_t } \Bigm\vert_q \delta^{\ell} = -  I^{\ell} \;\;,
& (25a)$^{\ell}$ \cr
&{\partial^2 \over \partial^2_t } \Bigm\vert_q \delta^{\ell}
+ 2 {{\dot a} \over a} {\partial \over \partial_t } \Bigm\vert_q \delta^{\ell} -
4 \pi G \varrho_H \delta^{\ell} = 0 \;\;. &(25b)$^{\ell}$ \cr}
$$
These equations can be obtained by linearizing (25).
On the contrary, the Lagrangian linear solution solves the exact
equations (25) for $II = 0$.

I emphasize that the similarity between the linear and the
nonlinear case restricted to $II=0$ is nontrivial: a nice excercise is
to compute the equation for the conventional density contrast $\delta$
from the nonlinear equation
(25b) for $II=0$ (solved by the Lagrangian first--order
solution) using the
definitions $\Delta:=\delta / (1 + \delta)$ and $\; \dot { }:=\partial_t
|_q + {1\over a}\vec u \cdot \nabla_q$, and to compare with the
Eulerian linear equation (25b)$^{\ell}$:
$$
\ddot \delta \;+\; 2 {\dot a \over a} \; \dot \delta \;-\;
4 \pi G \rho_H \delta \;+
$$
$$
{\ddot \delta} \delta \;-\; 2 {\dot \delta}^2 \;+\;
2 {\dot a \over a} \; {\dot \delta} \delta \;-\;
4 \pi G \rho_H \delta^2 \;=\;0\;\;\;.\eqno(26)
$$
Hence, first--order Lagrangian perturbations
involve nonlinearities
due to nonlinearities in the dependent variable $\delta$, but also
due to products of $\vec u$ and $\delta$ arising from the convective
derivative (contained in the overdot).

This exercise demonstrates the inherently
nonlinear character of a Lagrangian perturbation approach.
Essentially, this property can be traced back to the implicit
solution of the Eulerian convection of the flow $(\vec u
\cdot \nabla_q)\vec
u$ by the Lagrangian time--derivative, and to the fact that the density is
exactly known for any solution. 

The second--order Lagrangian theory takes 
the term $II$ in equation (25) approximately into
account, and thus covers essential effects of the tidal action on the fluid.

\subsection{Limitations of the Lagrangian perturbation theory}

In order to recognize the limitations of the Lagrangian perturbation approach,
two classes of arguments can be given: the first is analytical, based on the 
approach itself; the second is furnished by comparisons with N--body simulations.
Since the latter is the topic of another lecture (by Peter Coles, and ref. therein), 
I here concentrate on the former, and only give two illustrations of the latter.

Strictly speaking, the Lagrangian perturbation approach is limited by construction to 
the regime $t< t_c$,, where 
$$
|p_{i\1 j} (t_c)|<< a(t_c) \;\;.\eqno(27)
$$
This condition is very conservative in view of the
success of these approximations if followed up to shell--crossing, i.e., $|p_{i,j} (t_c)|
= {\cal O}(a(t_c))$ 
and beyond (see Coles' lecture). 
The condition (27) has its counterpart in the condition 
$\delta << 1$ in Eulerian perturbation theory.
At the epoch of shell--crossing, all orders of the perturbations yield 
displacements which are of comparable magnitude; consequently the perturbation approach 
breaks down there (we only expect it to converge before shell--crossing). 
In practice, however, we want to follow the solutions beyond 
shell--crossing (the Lagrange--Newton system is still regular in this regime,
i.e., the individual trajectories don't feel their intersection with others).
This demands that we must argue whether and how this extrapolation is justified.
The examples given in the previous subsection suggest that the extrapolation
can indeed be meaningful.

As mentioned earlier, in the regime after shell--crossing $J<0$, the mapping $\vec f$
is no longer unique and the inverse $\vec h$ is multivalued. 
If we assume that the density can be written as a superposition of the partial densities
corresponding to the different streams, then a reasonable extension into the multi--stream
regime for the density is  
$$
\varrho \lbrack{\vec x},t\rbrack = \sum_{{\gamma}=1}^{n} \varrho({\vec h}_{\gamma}
({\vec x},t),t)
= \sum_{{\gamma}=1}^{n} {\ueber{\varrho}{o} ({\vec h}_{\gamma} ({\vec x},t))
\over \vert\det
(\, f_{i\1 k}({\vec h}_{\gamma} ({\vec x},t),t)\,)\vert} \;\;;
\eqno(28)
$$
the total density in the $n$--stream region is the sum of the moduli of the
individual densities of the streams.
This extension is implicitly used when we construct the density field from displaced particles:
the formula (28) is approximated by the collection of particles in an Eulerian grid
at times after shell--crossing.
However, due to the presence of self--gravity the streams interact 
and the formula (28) 
has to be taken with caution. Indeed, expressing the superposition assumption 
in terms of the accelerations we encounter a problem:
In the multi--stream region there are $n$ fluid elements at the same Eulerian 
position, thus the total gravitational field strength at this position would
have to be added up similar to (28) and the fluid particle should be pushed by this
total strength. On the contrary, the Lagrangian solutions still push the fluid elements 
with their individual accelerations. In other words, Einstein's equivalence principle 
of inertial and gravitational mass is violated: the particles are not accelerated by the 
total gravitational field strength. 
We have to find extensions of the theory presented (e.g., in the framework of the 
Vlasov--Poisson system) to properly deal with this problem.
  
\bigskip

The performance of the Lagrangian perturbation solutions after shell--crossing
as inferred from a number of comparisons with N--body simulations appears to be weak 
due to the shortcomings discussed above. This is most easily visible in 
pancake models $\lbrack 13 \rbrack$.
If we consider hierarchical
cosmogonies, especially models which involve much small--scale 
power, then the power spectrum of fluctuations must be sufficiently steep to avoid 
shell--crossing on small scales. Otherwise, 
these high--frequency components have to be truncated or
filtered (see Coles lecture and ref. therein).
In the former case, a spectral index $n=-3$ in a power law spectrum $\propto 
\vert \vec k \vert^n$ lies close to the limit between ``pancake--type'' and 
``hierarchical'' cosmogonies having $n>-3$. This is so, because in the 
case $n=-3$ the integrated power per wave number interval is approximately constant
(logarithmic), so that all waves enter the nonlinear regime at about the same 
time (Fig.5). For more small--scale power the smoothing scale 
of high--frequency modes has to be close to (but 
smaller than) the nonlinearity scale. In Fig.6, I present the result of a comparison 
for a variant of the CDM cosmogony, the BSI model (see Gottl\"ober's lecture, 
this volume) with less small--scale power than standard CDM.  

\begin{figure}

\psfig{figure=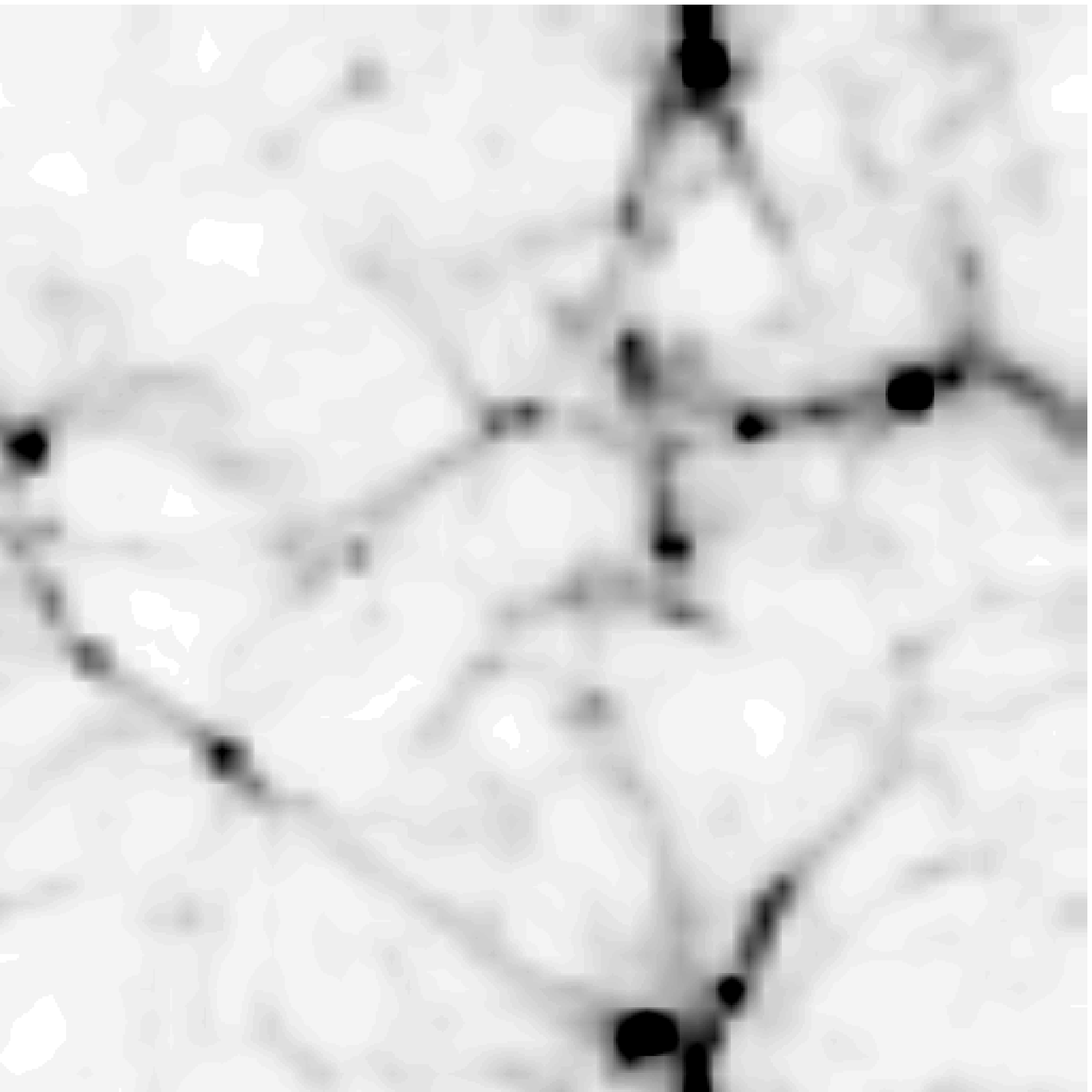,height=7.0cm,width=7.0cm}
\vskip -7.1 true cm
\rightline{\psfig{figure=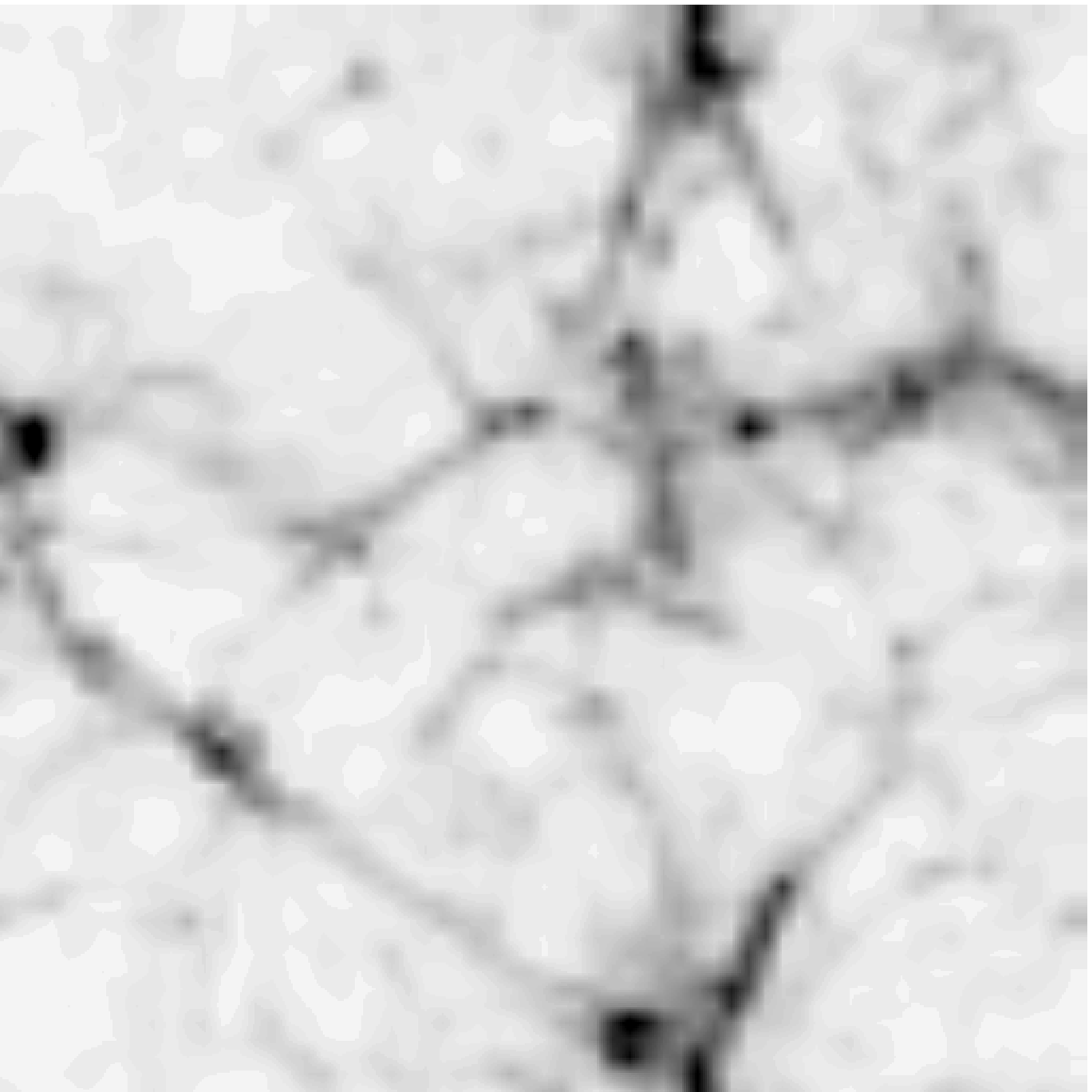,height=7.0cm,width=7.0cm}}

\caption{
A slice of the density field is shown for a PM simulation (left) and the 
second--order Lagrangian approximation 
(right panel) (taken from $\lbrack 27 \rbrack$).
The spectrum of fluctuations was given by a powerlaw with index $n=-3$, which can be taken 
to discriminate between the ``pancake'' and ``hierarchical'' 
regimes of structure formation.}

\end{figure}

\begin{figure}

\psfig{figure=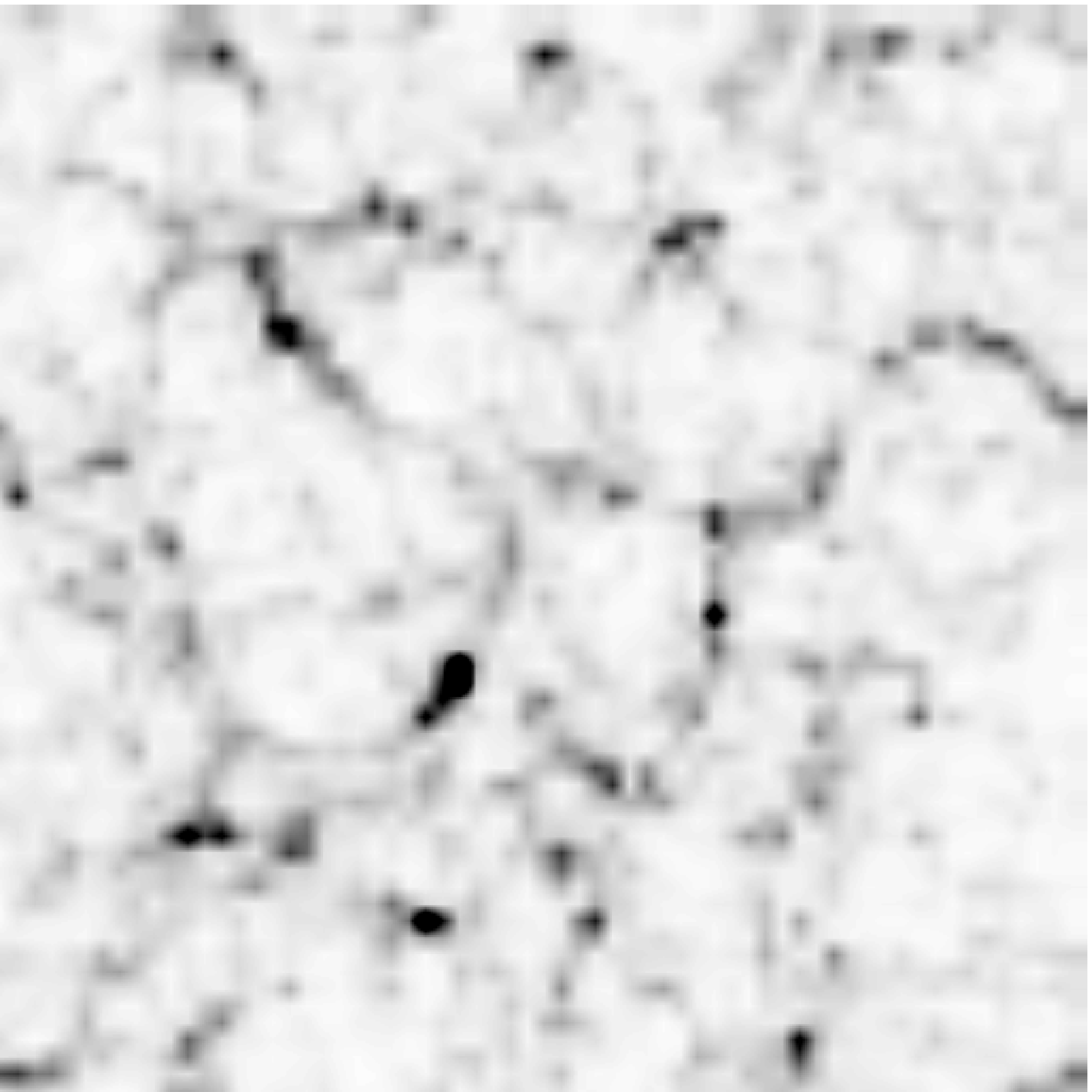,height=7.0cm,width=7.0cm}
\vskip -7.1 true cm
\rightline{\psfig{figure=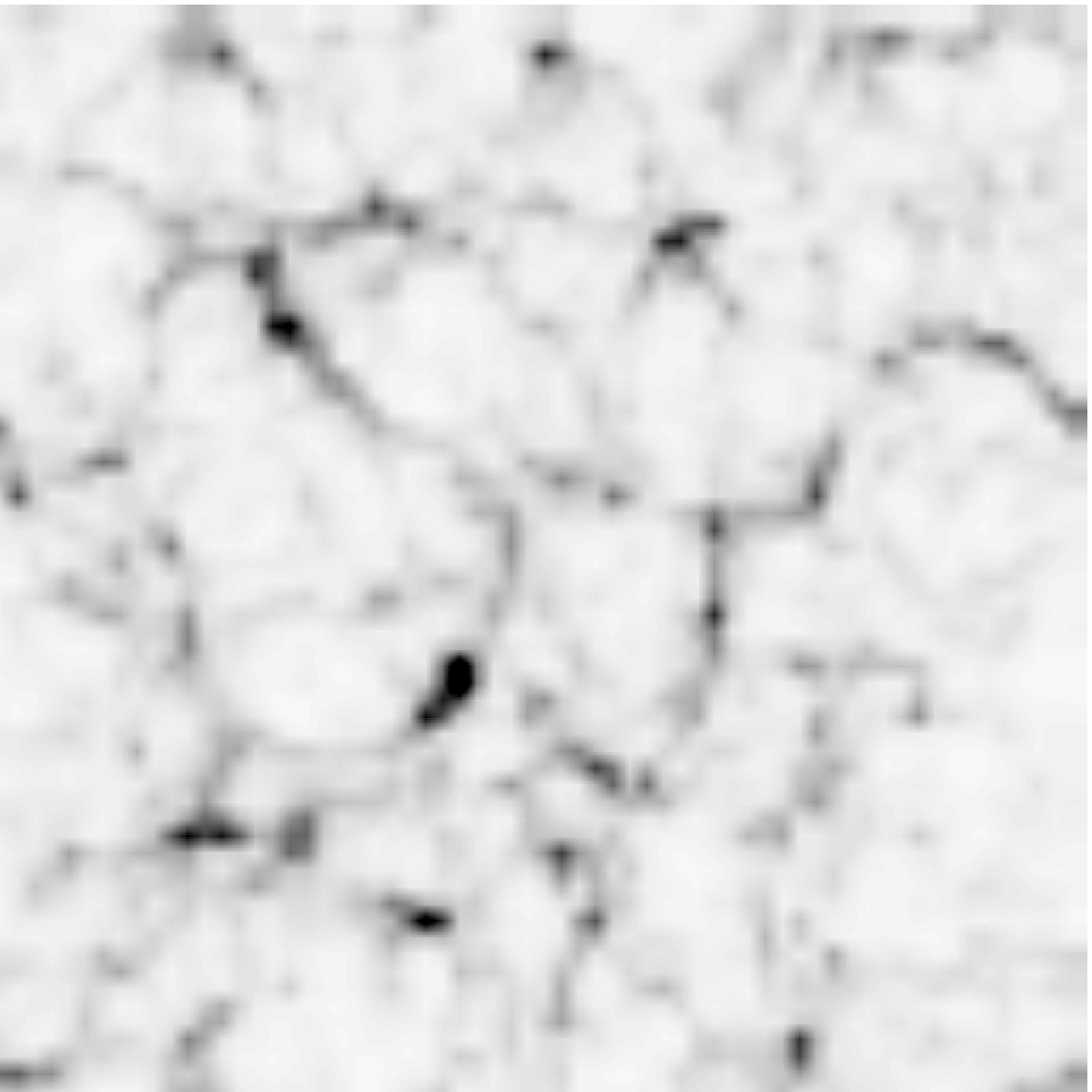,height=7.0cm,width=7.0cm}}

\caption{
A slice of the density field 
in a $200$h$^{-1}$Mpc box is shown for a PM simulation (left) and the 
second--order Lagrangian approximation 
(right panel) (taken from $\lbrack 33 \rbrack$).
The spectrum of fluctuations was a (COBE normalized)
non--standard CDM spectrum with broken scale invariance 
(BSI). The spectrum had to be smoothed at the high--frequency end for the Lagrangian 
scheme to avoid substantial 
post--singularity evolution.}

\end{figure}

\bigskip
\smallskip\noindent
Let us again summarize the main conclusions of this section:

\medskip
\noindent$\bullet$ In a Lagrangian perturbation approach only \underbar{one} variable, the 
trajectory field $\vec f$, is perturbed. The perturbation is evaluated along the 
perturbed orbits. Although the perturbations of the orbits might be small, the 
Eulerian fields evaluated along the perturbed orbits can experience large changes. 

\medskip
\noindent$\bullet$ The first--order Lagrangian approximation reduces to the 
\underbar{``Zel'dovich--approximation''}, if the initial data are restricted such that 
peculiar--velocity and peculiar--acceleration are parallel.

\medskip
\noindent$\bullet$ Imposing \underbar{periodic boundary conditions} 
on the cosmic peculiar--fields
is necessary to guarantee 1. \underbar{uniqueness} of the solutions, and 2.,
to assure that the \underbar{average flow} is given by the homogeneous ``background'' 
solutions. 

\medskip
\noindent$\bullet$ Higher--order effects amount to additional internal structures such
as \underbar{``second generation''} pancakes, filaments and clusters. They also accelerate
the collapse.

\medskip
\noindent$\bullet$ A Lagrangian perturbation solution inherently includes 
nonlinear terms.
This is also true for the first--order solution, which is linear in Lagrangian space.

\medskip
\noindent$\bullet$ The temporal limit of application of the Lagrangian schemes is reached
at the epoch of shell--crossing, where the displacements for each order attain similar
magnitudes. Following these schemes beyond shell--crossing time requires truncation of 
high--frequency modes in the initial fluctuation spectrum, thus defining a spatial 
lower limit of application, which roughly corresponds to galaxy group mass scales, 
$10^{13} M_{\odot}$ $\lbrack 26 \rbrack$, $\lbrack 16 \rbrack$.  

\bigskip\medskip\noindent
The Lagrangian perturbation approach is investigated to arbitrary order in 
$\lbrack 21 \rbrack$. 
In that paper as well as in the review articles  
$\lbrack 3 \rbrack$, $\lbrack 6 \rbrack$, $\lbrack 14 \rbrack$
and $\lbrack 30 \rbrack$ you may find references to related work.    
\vfill\eject
\noindent{\bf Acknowledgments:} This work considerably gained from many discussions
with J\"urgen Ehlers. I wish to thank him and H. Wagner for helpful 
comments on the manuscript. 
I acknowledge financial support by the ``Sonderforschungsbereich 375--95 f\"ur
Astro--Teilchenphysik'' der Deutschen Forschungsgemeinschaft.

\section*{References}

\ref$\lbrack 1 \rbrack$
Bertschinger E., Jain B., {\sl Ap.J.} {\bf 431} (1994) 486.
\ref$\lbrack 2 \rbrack$
Bertschinger E., Hamilton A.J.S., {\sl Ap.J.} {\bf 435} (1994) 1.
\ref$\lbrack 3 \rbrack$
Bertschinger E., in {\sl Cosmology and Large Scale Structure} 
Proc. Les Houches XV Summer School (1995), Elsevier Science Publishers B.V., 
in press.
\ref$\lbrack 4 \rbrack$
Bildhauer S., Buchert T., Kasai M., {\sl Astron. Astrophys.} {\bf 263} (1991) 23.
\ref$\lbrack 5 \rbrack$
Bouchet F.R., Juszkiewicz R., Colombi S., Pellat R., {\sl Ap.J.} {\bf 394} 
(1992) L5.
\ref$\lbrack 6 \rbrack$
Bouchet F.R., Colombi S., Hivon E., Juszkiewicz R., {\sl Astron. Astrophys.}
(1995) in press.
\ref$\lbrack 7 \rbrack$
Buchert T., G\"otz G., {\sl J. Math. Phys.} {\bf 28} (1987) 2714.
\ref$\lbrack 8 \rbrack$
Buchert T., {\sl Astron. Astrophys.} {\bf 223} (1989) 9.
\ref$\lbrack 9 \rbrack$
Buchert T., {\sl M.N.R.A.S.} {\bf 254} (1992) 729.
\ref$\lbrack 10 \rbrack$ 
Buchert T., {\sl Astron. Astrophys.} {\bf 267} (1993) L51.
\ref$\lbrack 11 \rbrack$
Buchert T., Ehlers J., {\sl M.N.R.A.S.} {\bf 264} (1993) 375.
\ref$\lbrack 12 \rbrack$ 
Buchert T., {\sl M.N.R.A.S.} {\bf 267} (1994) 811.
\ref$\lbrack 13 \rbrack$
Buchert T., Melott A.L., Wei{\ss} A.G., {\sl Astron. Astrophys.} 
{\bf 288} (1994) 349. 
\ref$\lbrack 14 \rbrack$
Buchert T., {\sl Phys. Rep.} (1995) submitted.
\ref$\lbrack 15 \rbrack$
Buchert T., Ehlers J., {\sl M.N.R.A.S.} (1995) to be submitted.
\ref$\lbrack 16 \rbrack$
Buchert T., Melott A.L., Wei{\ss} A.G., in {\sl 11th Potsdam Cosmology
Workshop on Large--scale Structure in the Universe}, eds.: M\"ucket J., 
Gottl\"ober S., M\"uller V. (1995), World Scientific, in press.
\ref$\lbrack 17 \rbrack$
Buchert T., Karakatsanis G., Klaffl R., Schiller P., Wei{\ss} A.G.,  
{\sl Astron. Astrophys.} (1995) to be submitted.
\ref$\lbrack 18 \rbrack$
Catelan P., {\sl M.N.R.A.S.} (1995) in press.
\ref$\lbrack 19 \rbrack$
Doroshkevich A.G., Kotok E.V., Novikov I.D., 
Polyudov A.N., Shandarin S.F., Sigov Yu.S., {\sl M.N.R.A.S.} {\bf 192} (1980) 321.
\ref$\lbrack 20 \rbrack$
Ehlers J., {\sl Akad. Wiss. Lit. Mainz, Abh. Math.--Nat. Klasse
11} (1961) p.793 (in German); translated (1993): {\sl G.R.G.} {\bf 25} 1225.
\ref$\lbrack 21 \rbrack$
Ehlers J., Buchert T., (1995) in preparation.
\ref$\lbrack 22 \rbrack$
Ellis G.F.R., in {\sl General Relativity and Cosmology}, ed. by
R. Sachs (1971), N.Y.: Academic Press.
\ref$\lbrack 23 \rbrack$
Ellis G.F.R., Dunsby P.K.S., {\sl Ap.J.} (1995) in press. 
\ref$\lbrack 24 \rbrack$
Kofman L., Pogosyan D., {\sl Ap.J.} {\bf 442} (1995) 30.
\ref$\lbrack 25 \rbrack$
Matarrese S., Pantano O., Saez D., {\sl M.N.R.A.S.} (1995) in press.
\ref$\lbrack 26 \rbrack$
Melott A.L., {\sl Ap.J.} {\bf 426} (1994) L19.     
\ref$\lbrack 27 \rbrack$
Melott A.L., Buchert T., Wei{\ss} A.G., {\sl Astron. Astrophys.} 
{\bf 294} (1995) 345.
\ref$\lbrack 28 \rbrack$
Peebles P.J.E., {\sl The Large Scale Structure of the Universe} (1980) 
Princeton Univ. Press.
\ref$\lbrack 29 \rbrack$
Peebles P.J.E., {\sl Ap.J.} {\bf 317} (1987) 576.
\ref$\lbrack 30 \rbrack$
Sahni V., Coles P., {\sl Phys. Rep.} (1995) in press.
\ref$\lbrack 31 \rbrack$
Serrin J., in {\sl Encyclopedia of Physics} {\bf VIII.1} (1959) Springer.  
\ref$\lbrack 32 \rbrack$
Szekeres P., Rankin J.R., {\sl J. Austral. Math. Soc.} {\bf 20(B)} (1977) 114.
\ref$\lbrack 33 \rbrack$
Wei{\ss} A.G., Gottl\"ober S., Buchert T., {\sl M.N.R.A.S.} (1995) 
in press.  
\ref$\lbrack 34 \rbrack$
Zel'dovich Ya.B., {\sl Astron. Astrophys.} {\bf 5} (1970) 84.
\ref$\lbrack 35 \rbrack$
Zel'dovich Ya.B., {\sl Astrophysics} {\bf 6} (1973) 164.

\end{document}